\documentclass[12pt]{article}
\usepackage[authoryear]{natbib}
\usepackage{amsmath,amssymb,amsfonts,amsthm}
\usepackage{epic,eepic,epsfig,longtable}
\usepackage{multirow,verbatim}
\usepackage{array}
\usepackage{comment}

\usepackage{epsfig}
\usepackage{setspace}
\usepackage{color}

\makeatletter
\def\singlespace{\def\baselinestretch{1}\@normalsize}

\makeatletter
\def\singlespace{\def\baselinestretch{1}\@normalsize}

\numberwithin{equation}{section}

\newcommand{\bfm}[1]{\ensuremath{\mathbf{#1}}}

\def\ba{\bfm a}   \def\bA{\bfm A}  
\def\bb{\bfm b}     
\def\bc{\bfm c}     
     
   \def\bE{\bfm E}  \def\EE{\mathbb{E}}
\def\bff{\bfm f}  \def\bF{\bfm F}  
\def\bg{\bfm g}   \def\bG{\bfm G}  
     
   \def\bI{\bfm I}

   \def\bM{\bfm M}

   \def\bP{\bfm P}  
     
\def\br{\bfm r}   \def\bR{\bfm R}  \def\RR{\mathbb{R}}
\def\bs{\bfm s}     
     
   \def\bU{\bfm U}  
\def\bv{\bfm v}   \def\bV{\bfm V}  
\def\bw{\bfm w}   \def\bW{\bfm W}  
\def\bx{\bfm x}   \def\bX{\bfm X}  
   \def\bY{\bfm Y}  
\def\bz{\bfm z}   \def\bZ{\bfm Z}

\def\calF{{\cal  F}}

\newcommand{\bfsym}[1]{\ensuremath{\boldsymbol{#1}}}

 \def\balpha{\bfsym \alpha}
 \def\bbeta{\bfsym \beta}			 \def\bPhi{\bfsym \Phi}
 \def\bgamma{\bfsym \gamma}             \def\bGamma{\bfsym \Gamma}
 \def\bdelta{\bfsym {\delta}}

 \def\btheta{\bfsym {\theta}}           
           
 \def\bsigma{\bfsym \sigma}             \def\bSigma{\bfsym \Sigma}
         
           \def\bOmega {\bfsym {\Omega}}
 \def\brho   {\bfsym {\rho}}      \def\bpsi {\bfsym {\psi}}			\def\bphi{\bfsym \phi}
 			\def\bUpsilon{\bfsym \Upsilon}
 \def\bxi{\bfsym {\xi}}

\DeclareMathOperator{\latent}{latent}

\DeclareMathOperator{\Cov}{Cov}

\DeclareMathOperator{\diag}{diag}
\DeclareMathOperator{\E}{\mathbb E}

\DeclareMathOperator{\sgn}{sgn}

\DeclareMathOperator{\Var}{Var}

\def\newpage{\vfill\eject}

\def\vecc{\mbox{vec}}

\def\today{\ifcase\month\or
  January\or February\or March\or April\or May\or June\or
  July\or August\or September\or October\or November\or December\fi
  \space\number\day, \number\year}
\def\Cov{\mbox{Cov}}

\newdimen\biblioindent    \biblioindent=30pt

 at 8truept

\def\sgn{\mbox{sgn}}
\def\ConStable{\overset{\mathcal{L}\text{-}s}\longrightarrow}

\newcommand{\beq}{\begin{equation}}
  \newcommand{\eeq}{\end{equation}}
\newcommand{\beqn}{\begin{eqnarray}}
  \newcommand{\eeqn}{\end{eqnarray}}
\newcommand{\beqnn}{\begin{eqnarray*}}
  \newcommand{\eeqnn}{\end{eqnarray*}}

\def\BIAS{{\rm BIAS}}

\allowdisplaybreaks
\usepackage{verbatim}
\renewcommand{\baselinestretch}{1.66}
\numberwithin{equation}{section}
\theoremstyle{plain}
\newtheorem{thm}{Theorem}[section]

\newtheorem{ass}{Assumption}[section]
\theoremstyle{definition}
\newtheorem{exm}{Example}[section]
\newtheorem{rem}{Remark}[section]
\newtheorem{remark}{Remark}[section]
\newtheorem{algo}{Algorithm}[section]

\newcounter{CondCounter}

\def \bbP      {\mathbb{P}}

\usepackage{xr}
\begin{document}

\title{Uniform Inference for Characteristic Effects  of Large Continuous-Time Linear Models\footnote{ We are grateful to Torben Andersen, Donald Andrews,  Xiaohong Chen,  Federico Bandi, Robert Korajczyk, Viktor Todorov,   Olivier  Scaillet   for many valuable comments. We also thank the comments from the audience of the 2017 MEG workshop, the 12th Greater NY   Colloquium, 2018 Econometric Society Summer Meeting, 2018 Market Microstructure and High Frequency Data, 2018 Big data on financial markets,  and seminar participants at SUFE, CUHK, UPenn, Northwestern and Yale.}}
\author{
Yuan Liao 
\and Xiye Yang\thanks{Address: 75 Hamilton St., New Brunswick, NJ 08901, USA. Email:
\texttt{yuan.liao@rutgers.edu}, \texttt{xiyeyang@econ.rutgers.edu}}
}


\maketitle

 \singlespacing

 \begin{abstract}

We consider continuous-time models with a large panel of moment conditions, where the structural parameter depends on a set of  characteristics, whose effects are of interest. The leading example is the linear factor model  in financial economics where factor betas depend on observed characteristics such as firm specific instruments and macroeconomic variables, and their effects  pick up long-run time-varying beta fluctuations. We  specify the  factor betas as the sum of  characteristic effects and an orthogonal idiosyncratic  parameter that captures high-frequency movements. It  is often the case that researchers do not  know whether or not the latter exists, or its strengths, and thus the inference about the characteristic effects  should be valid uniformly over  a  broad class of data generating processes  for idiosyncratic parameters. We construct our estimation and inference in a two-step continuous-time GMM framework.   It is found that   the limiting distribution of the estimated characteristic effects has a discontinuity when the variance  of the idiosyncratic parameter is near the boundary (zero), which makes   the usual ``plug-in" method using  the estimated asymptotic variance only  valid pointwise and may produce either over- or under- coveraging probabilities. 
 We show that  the uniformity can be achieved by  cross-sectional bootstrap. Our procedure allows both known and estimated factors, and    also features a bias correction for the effect of estimating unknown factors.  



 \end{abstract}



\newpage

\onehalfspacing

\section{Introduction}

Conditional factor models have been playing an important role in  capturing the time-varying sensitivities of individual outcomes to the  factors, in which the factor betas  are varying over time. 
In this paper, we  study a continuous-time conditional factor    model: 
\begin{gather}
\begin{split} \label{eq:Y}
    \bY_t =\bY_0 &+ \int_0^t \balpha_s ds + \int_0^t \bbeta_s d\bF_s + \bU_t
\end{split}
\end{gather}
where $\bY_0$ is the starting value of the outcome process   at time 0; the model contains a factor, idiosyncratic, and a drift process $\{(\bF_t, \bU_t, \balpha_t): t\geq 0\}$. Here $(\bY_t, \balpha_t, \bU_t)$ are all high-dimensional (whose dimension $p\to\infty$), and the dimension of the factor process $\bF_t$ is fixed. 
 In addition, the process $\{\bbeta_t: t\geq 0\}$ represents the factor loadings (or ``betas"), which is assumed to be stochastic in this paper.  Model (\ref{eq:Y}) covers  many uses of factor models in asset pricing.  
  In financial economics,  extensive empirical studies have shown that assets' individual betas can be largely explained by  their ``characteristics" (or called ``instruments").  These include lagged characteristics  that are common to all stocks,   characteristics specific to individual stocks, as well as observations of other   firm instruments \citep{gagliardini2016time}.   Estimated betas as functions of the conditioning  characteristics represent the effects of characteristics   on firm specific sensitivities to the risk factors.  They pick up   long-run  patterns and fluctuations in the betas.  
 Therefore, estimating the characteristic effects on the individual betas is one of the central econometric tasks in  financial economics.   Given the importance of the topic,  the  econometric problem, however, is   challenging for the reasons we shall elucidate  below. 

 Let $\bbeta_{lt}$ denote the $K$-dimensional factor betas  of the $l$-th individual at time $t$. Let $\bX_{lt} $ be a vector of  observed characteristics that may be varying across individuals and times, and $\bX_t=\{\bX_{lt}: l\leq p\}$.
 We model: for each $l=1,...,p$,
\begin{eqnarray}\label{e1.1}
\bbeta_{lt} &=&\bg_{t}(\bX_{lt}) +\bgamma_{lt},\quad  \mathbb E(\bgamma_{lt}| \bX_{t}  )=0, 
\end{eqnarray}
where  $\bg_t(\cdot)$ is an unknown time-varying nonparametric function that is assumed to be well approximable by a    sieve representation. Here, each component $\gamma_{lt,r}$ of $\bgamma_{lt}$ ($r\leq K$) satisfies, for some constant $C>0$,
\begin{eqnarray}\label{e1.1add}
\Var(\gamma_{lt,r}| \bX_{t}  )\in[0, C] \text{ almost surely}. 
\end{eqnarray}
The main message from   (\ref{e1.1})-(\ref{e1.1add}) is that  in the identification condition $\bg_{t}(\bX_{lt})= \mathbb E( \bbeta_{lt}| \bX_{lt}  )  $, the variance of the ``error components"
  $\bgamma_{lt}:= \bbeta_{lt} - \bg_{t}(\bX_{lt})    $  is defined on a compact set that includes zero, and can be either exactly on, arbitrarily close to, or bounded away from its ``boundary".
  Thus we allow    \textit{   arbitrarily unknown  signal strengths} of $\bgamma_{lt}$. 
So  the economic meaning of model    (\ref{e1.1}) is that  the factor beta is decomposed as  the sum of  two components: (i) a   nonparametric function of the observed instruments, $\bg_{lt}:=\bg_{t}(\bX_{lt})$, which we call ``characteristic betas",  and  (ii) a  time-varying and individual specific component, $\bgamma_{lt}$, which we call ``idiosyncratic betas".    
 The characteristic beta $\bg_{lt}$ picks up long-run beta patterns and fluctuations,  and possesses less volatile, 
    while $\bgamma_{lt}$  captures high frequency movements in beta,
and 
represents the remaining time-varying individual factor sensitivities after conditioning on the observed characteristics. In addition, the strength of the idiosyncratic betas is allowed to be arbitrary, reflecting the  nature that the explainability from the characteristics is unknown.

  The goal of this paper is to provide a \textit{uniformly valid}   inference of $\bg_{lt}$, the characteristic effects on factor betas.     By ``uniformly valid", we mean the coverage probability is asymptotically correct uniformly over a broad class of data generating processes (DGPs) that
 allows various possible signal strengths of $\bgamma_{lt}$, measured by a weighted cross-sectional variance:
   $$
   \bV_{\gamma, t} := \Var\left(\frac{1}{\sqrt{p}}\sum_{m=1}^ph_{t,ml}\bgamma_{mt}\bigg{|}\bX_t\right).
   $$   
We assume $\bgamma_{lt}$ to be conditionally cross-sectionally independent given $\bX_t$, and find that  the strength of $  \bV_{\gamma, t}$
      plays a crucial role in the asymptotic behavior of estimated characteristic effect, and affects both the rate of convergence and limiting distributions. 
      Here $h_{t,ml}$ is a function of $\bX_t$   whose definition will be clear in the paper. 
       In particular, the asymptotic distribution of $\bg_{lt}$ has a ``discontinuity" when the strength of $\bgamma_{lt}$, the eigenvalues of $ \bV_{\gamma, t}$,  are  near zero.    
    As a consequence,     the usual \textit{pointwise}  inference procedures  under a fixed DGP   
  only produce confidence intervals that are   valid for specific DGPs, therefore potentially produce misleading inferences.    Specifically, 
  benchmark methods  in the   literature, which  ignore the  high-frequency beta dynamics in $\bgamma_{lt}$,  would  produce \textit{under-coveraging} confidence intervals of the characteristic effects.            On the other hand, we show that even if $\bgamma_{lt}$ is allowed,  standard ``plug-in"  procedures using the estimated  asymptotic variances  do not produce uniformly correct coverage probabilities, because they  require very strong signal strengths of $\bV_{\gamma, t}$, leading to \textit{over-coveraging } confidence intervals when the signal strength is weak.   The discontinuity    issue here is similar to the problem of estimating parameters on a boundary. 
As is shown  by, e.g.,   \citep{andrews1999estimation} and \citep{ketz2017testing}, when a test statistic has a discontinuity in its limiting distribution, as occurs in estimating parameters on a boundary and in random coefficients models, pointwise asymptotics can be very misleading.

We reply on a cross-sectional bootstrap to achieve the uniform inference. It is important to note that the  employed bootstrap is cross-sectional,  which resamples  the cross-sectional  individuals  while  keeping all the serial observations for each resampled individual. This procedure is essential because the discontinuity arises when the   cross-sectional variance $   \bV_{\gamma, t}$ is near the boundary, and the cross-sectional bootstrap avoids estimating $   \bV_{\gamma, t}$ in the current context. We
 show that the bootstrap procedure leads to a  correct asymptotic coverage probability   and is uniformly valid over a large class of  DGPs,  and explain the reasons in detail.

  Allowing $   \bV_{\gamma, t} $ to be unrestricted on a compact set that includes zero as the boundary point  is the main motivation of this paper, which arises from the following practical consideration.
 While characteristics may fully explain the factor betas at times when they are just updated and made publicly available ($\bgamma_{lt}\approx0$), there are also   times when factor betas contain  either unmeasurable or  high-frequency components  that  are more volatile and cannot be captured by the characteristics ($\bgamma_{lt} $ very different from zero). 
 In those occasions,   modeling the beta  as fully specified functions of observed characteristics can be   very restrictive. This is particularly true  for high-dimensional and high-frequency factor models in empirical asset pricing, 
 where individual factor betas demonstrate   large heterogeneity when the number of assets is large, and  assets' returns  are available at a very  high frequency. 
 On the contrary,    characteristics  such as the firm sizes and book-market values, often vary more smoothly and are measured at a much lower frequency, often (but not always) leaving large portions of  stock betas' dynamics  unexplained.    
As we show in this paper, without taking  into account the  high-frequency movements of betas after conditioning on the characteristics,   the inference procedures of characteristics' effects   are  not asymptotically valid. Unfortunately, this is often  the case in  the financial economic  literature, which has been dominated by modeling betas as fully specified functions of the observed characteristics, including both parametric (e.g., \cite{shanken1990intertemporal,  cochrane1996cross, ferson1999conditioning, avramov2006asset,  gagliardini2016time}) and nonparametric models, e.g.,  \cite{CL07}  and \cite{CMO}.
   As is shown by \cite{ghysels1998stable},  misspecifying beta risk may result in serious pricing errors that might even be larger than those produced by an unconditional asset pricing model.

 \subsection{Many moment conditions in continuous-time}
 
 The model can be generalized to the high-dimensional continuous-time linear moment conditions framework.  For each $l\leq p,$  the parameter  $  \bbeta_{lt} $   is identified by the following moment condition:
 \begin{equation}\label{e1.3}
  \Psi_l(\bbeta_{lt},  \bc_{z, lt})  =0,\quad \forall t\in[0, T]
 \end{equation}
where   $\bc_{z,lt}=d[\bZ_l,\bZ_l]_t/dt$ is  the instantaneous quadratic variation process of $\bZ_l=\{\bZ_{lt}\}_{t\geq 0}$, 
and  $ \Psi_l(\bbeta,\bc)$ is a known   function  linear with respect to $\bbeta$. In addition,   we observe  a set of    time-varying characteristics  
 $\bX_{lt}$, so that we have  the following decomposition:
\begin{align} \label{eq1.4}
   \bbeta_{lt} = \bg_{t}(\bX_{lt}) + \bgamma_{lt}, \quad l=1,\cdots,p,
\end{align}
where $\mathbb E(\bgamma_{lt}|\bX_{t})=0$.   The model admits the linear factor model as a special case by setting   $\bZ_{lt}=(Y_{lt}, \bF_t)$  and 
 $$\Psi_l(\bbeta_{lt}, \bc_{z, lt})=\bc_{FF,t}\bbeta_{lt}  -\bc_{YF,lt},\quad \bc_{z, mt}=(\bc_{FF,t}, \bc_{YF,mt}). $$
  Here $\bc_{FF,t} $   and $\bc_{YF,lt} $ are  the quadratic covariances for the  processes   $\bF_t$ and $Y_{lt} $.
Moreover, the above model also admits a continuous-time linear regression model with individual-specific regressors: e.g., \cite{barndorff2004econometric,mykland2006anova,kalnina2012nonparametric}, as well as the idiosyncratic volatility model applied by  \cite{ang2009high,herskovic2016common, li2016inference}.

We provide a general two-step  estimation procedure to make uniform inferences about the characteristic effect $\bg_{lt}= \bg_{t}(\bX_{lt}) $ for each fixed $l\leq p$. In step (i), we  estimate $\bbeta_{lt}$ by either directly solving  (\ref{e1.3}) or 
using generalized method of moments (GMM),    with the sample quadratic variation in place of $ \bc_{z, lt}$, and in step (ii), we estimate $\bg_{lt}$ by a standard nonparametric sieve regression on (\ref{eq1.4}) with the estimated $\bbeta_{lt}$.  We aim to construct a confidence interval $CI_{\tau}$ for $\bg_{lt} $ for any specific $l\leq p$ and at a specific time $t$, so that at the nominal level $1-\tau$,
$$
\lim_{p, T\to\infty} \sup_{\mathbb P\in\mathcal P}\left|\mathbb P\left(\bg_{lt}\in CI_{\tau} \right)-(1-\tau)\right|=0
$$
Here the probability measure $\mathbb P$ is taken uniformly over a broad  DGP class $\mathcal P$, which admits various 
 cross-sectional variations in $(\bgamma_{lt}$, $\bg_{lt})$ and  dynamics if they are also time-varying.
While our framework belongs to a more general class of two-step  GMM estimators, we encounter   a new feature as in the linear factor model we described earlier:  the  estimated $\bg_{lt}$ possesses a   discontinuity  on its limiting distribution because the strengths of $ \bV_{\gamma, t}  $ is unknown and can be near the boundary. This brings new challenges to achieving the uniformity in the inference.
  Uniformity in the above sense is essential  in this context,  because it makes the inference valid and robust to the unknown degrees of dynamics in $\bbeta_{lt}$, especially, the strengths of the cross-sectional variations in $\bgamma_{lt}$. 
 

 In the presence of ``boundary parameters", the asymptotic inference becomes nonstandard, and modified inference procedures have been proposed in both time series and cross-sectional models, e.g., \cite{ketz2017testing, ketz2018subvector,pedersen2017inference}. The main difference between our model  and those considered in the literature is that we do not model  $   \bV_{\gamma, t} $ explicitly as an unknown ``parameter", so it  does not appear  in the GMM objective function. In fact, the mapping from $\bgamma_{lt}$ to the limiting distribution of the estimated $\bg_{lt} $ is still continuous, which makes the cross-sectional bootstrap first-order valid.  We provide more detailed explanations in Section 2.

\subsection{The literature}
 
The   linear factor model covers  many useful models  in the arbitrage pricing theory and  models of linkages between international stock markets, as well as macroeconomics. Earlier literature include, e.g., \cite{CR, CK93,king1994volatility, SW02,BN02}. 
  With the massive use of the newly available datasets of intraday asset prices and large number of cross-sectional data, the continuous-time factor model of high dimensions have received extensive attentions in the recent high-frequency literature, such as \cite{Pelger:2016, ait2017using,fan2016incorporating, li2018jump}. 

The study of the effects of characteristics on betas   is an essential subject in financial economics.  
For instance, it is commonly known that firm  sensitivities to risk factors  depend on the firm specific raw size  and value characteristics.    As is noted by \cite{daniel1997evidence}, ``It is the firms' characteristics (size and ratios) rather than the covariance structure of returns that appear to explain the cross sectional variation in stock returns." 
 \cite{ang2012testing} also found that the market risk premium is less correlated with value stocks' beta (stocks with high book-to-market ratio) than with growth stocks' beta. 
Firms' momentum   is also one of the commonly used characteristics,  whose effect on the factor sensitivities has been found to be linearly growing with the momentum, indicating a constant effect. In addition,  \cite{ferson1999conditioning} found that  the lagged characteristics track variations in expected returns that is not captured by the Fama-French \citep{FF} three-factor model, and that these characteristics  have explanatory power on the factor loadings because they pick up betas' time-variation.
 In addition, the effect  of common  characteristics such as  the term spread and default spread 
 demonstrate significantly different volatiles among betas of individual stocks and portfolios, explaining the larger heterogeneity of the factor loadings for the former. Other empirical evidence that systematic risk is related to firm characteristics and business cycle variables is provided by \cite{jagannathan1996conditional, lettau2001resurrecting}, among many others.   

While most of the aforementioned works assume that the betas are fully explained by observed characteristics, 
 a similar decomposition to (\ref{e1.1}) was given by \cite{kelly2017instrumented},  where betas are  decomposed into a linear function of lagged  characteristics as well as an  unobservable  loading component.  They specifically require $\bgamma_{lt}$ to be ``strong", with cross-sectional variances that are bounded away from zero. 
 \cite{fan2016projected} and \cite{kim2018arbitrage} respectively studied  a model whose betas have a similar decomposition. They   did not study the   inference problem. 
  Our paper is also related to the continuous-time GMM framework of \cite{li2016generalized},  
but  is     different on the inference aspect, where we study the continuous-GMM estimation in the presence of high-dimensional linear moment conditions, and when combined with the nonparametric regression, there is  a  discontinuity issue  on the limiting distribution for the characteristic effect.   Other literature on continuous-time regression models can be found from \cite{barndorff2004econometric,mykland2006anova,  li2017adaptive}, among others. Also note that we  focus on the continuous components of the factor models, and factor model for the jump components was studied by  \cite{li2018jump}.


 The rest of this paper is organized as follows.  Section \ref{sec:uniform} informally discusses the issue of uniformity and the intuitive solutions using the cross-sectional bootstrap.  Section \ref{sec: model} describes the continuous-time conditional factor model driven by   stochastic processes.   We separately study the known and the unknown factor cases, and present the asymptotic results of the estimators.  Section \ref{s:gmm} extends the model to the more general continuous-time GMM framework, with many moment conditions.
  Section \ref{sec: realdata} presents real data applications on the high-frequency stock return data of firms from S\&P500.
Finally,   in the supplement, we give simulated example to examine the  uniformity of the proposed inference in finite sample, additional empirical findings,  as well as all the technical proofs.   

 \textbf{Notation:} We observe  data  every $\Delta_n$ unit of time and let $\Delta_n$ go  to zero in the limit. For any process $Z$, let $ \Delta_i^n Z=Z_{i\Delta_n}- Z_{(i-1)\Delta_n} = \int_{(i-1)\Delta_n}^{i\Delta_n} dZ_t$. For simplicity, we will denote $Z_{i\Delta_n}$ by $Z_i$.
 We use the symbol $\ConStable$ to denote stable convergence in law. We say a constant a \textit{absolute constant} if it does not depend on any pointwise DGP. Let $\bI_d$ be the $d\times d$ dimensional identity matrix. For a matrix $\bA$, we use $\lambda_{\min}(\bA)$ and $\lambda_{\max}(\bA)$ to respectively denote its smallest and largest eigenvalues. In addition, let $\|\bA\|:=\lambda_{\max}^{1/2}(\bA'\bA)$, and $\|\bA\|_\infty=\max_{ij}|(\bA)_{ij}|$. In addition,  we shall achieve   inferences  uniformly valid over  a large class of data generating process $\mathcal P$.   For a random sequence $X_n$, we write $X_n\asymp O_P(a_n)$ if $X_n=O_P(a_n)$ and $a_n/X_n=O_P(1)$.

 \section{Heuristic Discussions on the Uniformity Issue}\label{sec:uniform}
 
 \subsection{A discrete-time unconditional model}
While this paper studies continuous-time models, in this section we heuristically discuss   the issue we encounter on  the uniform inference using a discrete-time, unconditional factor model with observed factors. 
Consider the following discrete-time one-factor model:
\begin{equation}\label{e2.1}
y_{mt}= [\bx_{m}'\btheta+\bgamma_m]\bff_t+u_{mt},\quad m=1,...,p; \quad t=1,...,T.
\end{equation}
Here $(y_{mt}, \bff_t,\bx_m)$ are observable and of finite dimensions, in particular for ease of presentation, $\dim(\bff_t)=\dim(\bgamma_m)=1$.  The goal is to  make inference about $\btheta$, which is a $d\times 1$ vector. 
 We assume that  $\mathbb E(\bgamma_m|\bX)=0$,  $\mathbb E(u_{mt}|\bX, \bF, \bGamma)=0$, where $(\bX,\bGamma)=\{(\bx_m,\bgamma_m): m\leq p\}$ and $\bF=\{\bff_t: t\leq T\}$, and that $\{u_{mt},\bgamma_m\}$ are cross-sectionally independent across $m\leq p.$

A natural estimator for $\btheta$ is based on  a combination of cross-sectional and time-series regression:   
$$
\widehat \btheta=  \bs_f^{-1} \bs_x^{-1} \frac{1}{pT}\sum_{t=1}^T \sum_{m=1}^p\bx_m\bff_t y_{mt}  , 
$$
where   $ \bs_{f}=\frac{1}{T}\sum_{t=1}^T\bff_t^2$,  and   $\bs_x=\frac{1}{p}\sum_{m=1}^p\bx_m\bx_m'$. 
Then   $\widehat\btheta$ has the following   expansion
\begin{equation} \label{e2.2new}
\widehat\btheta-\btheta= \underbrace{  \bs_x^{-1}\frac{1}{p} \sum_{m=1}^p \bx_m \bgamma_m   }_{(\bb)}+ \underbrace{  \bs_f^{-1} \bs_x^{-1} \frac{1}{pT}\sum_{t=1}^T \sum_{m=1}^p\bx_m\bff_t u_{mt} }_{(\ba)}  .\end{equation} 
Thus the asymptotic distribution depends on the interplay of two leading terms. 
Term $(\bb)$ arises from the cross-sectional estimation, which has a rate $O_P(p^{-1/2}\|\bV_{\gamma}\|^{1/2})$, with   \begin{equation}\label{e2.3}
\bV_{\gamma}=\bs_x^{-1}\frac{1}{p}\sum_{m=1}^p \bx_m\bx_m'\Var(\bgamma_m|\bX)\bs_x^{-1},
\end{equation}
   where $\Var(\cdot | \bX)$ denotes the conditional  variance  given $\bX$.  Because  $\bgamma_m$'s are cross-sectionally independent so this term admits a cross-sectional central limit theorem (CLT). In addition, term $(\ba)$  has a rate $O_P((Tp)^{-1/2})$ because $u_{mt}$'s are conditionally independent across $(i,t)$. 

The first question to address   is, what are the final rate of convergence and the  limiting distribution? The key to this question is that  the strengths of the eigenvalues of $\bV_{\gamma}$ are unknown, and are arbitrarily supported on a compact set $[0, C]$ for $C>0.$
 If  $\bV_\gamma$ is \textit{weak and near boundary} (zero), whose  eigenvalues, treated as  sequences, decay at rate faster than $O_P(T^{-1})$,  then   $(\ba)$ is the dominating term, leading  to, for some covariance $\bv_1$,
  $$
 \sqrt{Tp} (\widehat \btheta-\btheta)\to^d \mathcal N(0, \bv_1),
  $$
whose asymptotic distribution and $\bv_1$ are  determined by $(\ba)$.
  Intuitively, this occurs when   the observed characteristics capture  almost all the beta fluctuations, leading to a fast rate of convergence.  On the other hand,  
 if  $\bV_\gamma$ is \textit{strong} with all eigenvalues bounded away from zero, $(\bb)$ becomes the dominating term, and  we simply have,  for $\bv_2:=\text{plim}_{p\to\infty}\bV_{\gamma}$, 
       $$ \sqrt{p}(\widehat \btheta-\btheta)
\to^d \mathcal N(0, \bv_2).
  $$
In this case, the  limiting distribution is determined by the cross-sectional CLT of $(\bb)$.   Intuitively, this  means when the idiosyncratic betas  have  strong cross-sectional variations, time series regression  is not helpful to remove their effects on estimating $\btheta$, and the cross-sectional projection  dominates.  This leads to a slower rate of convergence.

Consequently, there is a discontinuity on the limiting distribution of $\widehat \btheta-\btheta$ when $\bV_\gamma$ is near the boundary. In practice, anything in between the above two extreme cases might also happen, leading to an unknown rate of convergence $O_P(a_{pT})$, where $a_{pT}\in[(pT)^{-1/2}, p^{-1/2}]$.
This issue  is similar to the  problems in estimating     parameters that are possibly on the boundary of the parameter space  \citep{andrews1999estimation, andrews2010inference}. The problem arises as we do not pretest or know how strong $\bgamma$'s cross-sectional variation is, which can vary in a large class of data generating process.
Most of the financial econometric studies take the ``weak" case as the default assumption (e.g., \cite{ferson1999conditioning, gagliardini2016time, CMO}), while more recent studies   (e.g., \cite{fan2016projected, kelly2017instrumented, kim2018arbitrage}) provide evidence of the presence of the ``strong" case in some sampling periods. Above all, to our best knowledge, all the existing inferences are pointwise, and is not robust to the strength of $\bgamma$'s  variations. Pointwise inferences,   therefore, can be misleading. 

  \subsection{Drawbacks of the usual ``plug-in" method}
 
 The second question to address   is, what is the impact of the  unknown rate for $\bV_\gamma$ on the inference about $\btheta$? The ``standard" inference procedure is to plug-in the estimated asymptotic covariances for  $\bV_{\gamma}$ and $\bv_1$, using  their sample analogues. This procedure, however, works only pointwise, and does not provide a uniformly valid confidence interval.
 To understand the issue, consider the estimation of $\bV_{\gamma}.$   If $\bgamma_m$ were known, \cite{white1980heteroskedasticity}'s heteroskedastic  covariance estimator can be applied:
 \begin{equation}\label{ev:white}
 \widetilde{\bV}_{\gamma} =\bs_x^{-1}\frac{1}{p}\sum_{m=1}^p \bx_m\bx_m' \bgamma_m ^2 \bs_x^{-1} .
 \end{equation}
Replacing $\bgamma_m$ with its consistent estimator $\widehat\bgamma_m$, we obtain $\widehat\bV_{\gamma}= \bs_x^{-1}\frac{1}{p}\sum_{m=1}^p \bx_m\bx_m' \widehat\bgamma_m ^2 \bs_x^{-1}$. Then  $\widehat\bV_{\gamma}- \bV_{\gamma}$ has a decomposition
\begin{equation}\label{e2.2}
\underbrace{\bs_x^{-1}\frac{1}{p}\sum_{m=1}^p \bx_m\bx_m' [\widehat\bgamma_m^2-\bgamma_{m} ^2] \bs_x^{-1} }_{\gamma\text{-estimation error}}+\underbrace{\bs_x^{-1}\frac{1}{p}\sum_{m=1}^p\bx_m\bx_m' \left[\bgamma_m ^2-\Var(\bgamma_m|\bX)\right]\bs_x^{-1}}_{\text{LLN error}}
\end{equation}
where ``LLN error" refers to the error associated with the  law of large number.  The main issue is that the $\gamma$-estimation error cannot be uniformly controlled. 
Note that in the  ideal case  where $\btheta$ were known, one would estimate $\bgamma_m$ from (\ref{e2.1}) by running time series regression of $y_{mt}-\bx_m'\btheta\bff_t$ on $\bff_t$ for  each fixed $m\leq p$.  Then 
$\widehat\bgamma_m-\bgamma_m=\br_m$, where $\br_m:=\bs_{f}^{-1}\frac{1}{T}\sum_{t\leq T}\bff_tu_{mt}.$
This  leads to
$$
\gamma\text{-estimation error} \geq \bs_x^{-1}\frac{1}{p}\sum_{m=1}^p\bx_m\bx_m'\br_m^2  \bs_x^{-1}\asymp O_P( T^{-1}).
$$
This results in an estimation error $\|\widehat\bV_{\gamma}-\bV_{\gamma} \|$ being  {lower bounded} by an order $ O_P(  T^{-1})$, which is not negligible 
 whenever $\lambda_{\min}(\bV_{\gamma})=O_P(T^{-1})$ (corresponding to the case of  weak $\gamma$-signal).

 Consequently, the usual plug-in covariance estimator using $\widehat\bV_{\gamma}$ would lead to an asymptotically incorrect distribution, and over-coveraging probabilities. On the other hand, ignoring $\bV_{\gamma}$ would result in under-coveraging probabilities when  it  is  present. Hence it is    not uniformly valid.\footnote{More precisely, when $\bV_\gamma=O_P(T^{-1})$, plugging in its consistent estimator  over-estimates the asymptotic variance, leading to   valid but severely conservative inferences. }


 \subsection{The cross-sectional bootstrap}
 

To resolve  the uniformity issue, we propose to use the cross-sectional bootstrap, which is intuitive and very easy to implement. We simply take random samples with replacement across cross-sectional individuals $\{1,..., p\}.$ Once an individual  $l^*\in \{1,..., p\}$ is sampled, its associated entire time series $\{y_{l^*,t}\}_{t\leq T}$ is sampled.  Then we obtain the estimator $\widehat\btheta^*$ using the bootstrap data. Finally,  we  calculate the   critical value of $ \widehat \btheta^*-\widehat\btheta$ from a set of bootstrap estimators.   This procedure is very simple, but perhaps surprisingly,  leads to the desired uniform coverage for $\btheta$.

To prove the bootstrap validity, it is essential to show that 
  this procedure directly mimics the cross-sectional variations in $\{\bgamma_m\} $. To see this intuitively, we note that we can expand the bootstrap estimator as:   for  $\bR$ as an asymptotically negligible term,
  $$
  \widehat\btheta^*-  \widehat\btheta= {  \bs_x^{-1}\frac{1}{p} \sum_{m=1}^p (\bx_m ^*\bgamma_m ^*-   \bx_m \bgamma_m )  }+  {  \bs_f^{-1} \bs_x^{-1} \frac{1}{pT}\sum_{t=1}^T\bff_t  \sum_{m=1}^p(  \bx_m^* u^*_{it} -\bx_m u_{mt} )} 
  +\bR,
  $$
where $\{(\bx_m^*, \bgamma_m^*): i\leq p\}$ is a simple random sample from  $\{(\bx_m, \bgamma_m): i\leq p\}$ with replacement.
  Then  the bootstrap asymptotic variance of $\widehat\btheta^*$ is  analogously $ \frac{1}{Tp} \bv_1 +\frac{1}{p} \widetilde{ \bV}_{\gamma}$, where $\frac{1}{Tp}\bv_1$ is the asymptotic variance of term $(\ba)$ in (\ref{e2.2new}), and $\widetilde\bV_\gamma$ is defined in (\ref{ev:white}).   The only approximation error for   $ \bV_{\gamma}$ is therefore:
$$
\widetilde{ \bV}_{\gamma}- \bV_{\gamma}=\underbrace{\bs_x^{-1}\frac{1}{p}\sum_{m=1}^p\bx_m\bx_m' \left[\bgamma_m ^2-\Var(\bgamma_m|\bX)\right]\bs_x^{-1}}_{\text{LLN error}}.
$$
Consequently, the $\gamma$-estimation error component in (\ref{e2.2}) is avoided.  The LLN error is of a higher order than $\bV_{\gamma}$, regardless of the signal strength of $\bV_\gamma$.  For instance, suppose $\bgamma_m$ is generated from a rescaled sequence, that is, $\bgamma_m= b_{T}\bar\bgamma_m$, where $b_{T}\geq 0$ is a non-random arbitrary sequence, and $\bar\bgamma_m$ satisfies,  for $ C, c_0>0$, almost surely,  $$ \lambda_{\min}\left[\frac{1}{p}\sum_{m=1}^p\bx_m\bx_m'\Var(\bar\bgamma_m|\bX)\right]>c_0,\quad \mathbb E(\|\bar\bgamma_m\|^4|\bX)<C.$$Then the  LLN-error $=o_P(1)\bV_{\gamma}$. Hence the approximation error for the asymptotic variance of $\widehat\btheta$ is  negligible regardless of the  strength  $b_{T}$. The bootstrap validity can be achieved. 

\cite{andrews2000inconsistency} gave a generic counter-example showing that the usual bootstrap is inconsistent when the parameter is near the boundary of its space. So modified inference procedures have been proposed, e.g., see more recently, \cite{ketz2017testing, ketz2018subvector,pedersen2017inference}. 
 We note several important differences between our problem and that of the cited literature. The main difference is that   the mapping from the underlying DGP to the  asymptotic distribution of $\widehat \btheta$ is continuous in our setting. Such a mapping is  essentially $ \bGamma:\to \mathcal M(\bGamma)$,
$$
\mathcal M(\bGamma):=\underbrace{  \bs_x^{-1}\frac{1}{p} \sum_{m=1}^p \bx_m \bgamma_m   }_{(\bb)},
$$
where $\bGamma=\{\bgamma_m\}$.
  The main reason of achieving a   continuous mapping   is that  we \textit{ do not}  specify   $\Var(\bgamma_m|\bX)$  or $\bV_\gamma$ as  unknown parameters. 
All the parameters in our model, $\{(\btheta, \bgamma_m): i\leq p\}$, are inside the interior of their  parameter space $\Theta_\theta\times \otimes_{m=1}^p \Theta_\gamma$, where
$\Theta_\theta$ and $\Theta_\gamma$ are compact subsets respectively in  $\mathbb R^K$ and $\mathbb R$. 
Therefore when  estimating  $\btheta$, the loss function (e.g., least squares) does not explicitly depend on the unknown ``boundary parameter"  $\Var(\bgamma_m|\bX)$.  
In the absence of $\Var(\bgamma_m|\bX)$ in the loss function and with the continuous mapping $\mathcal M(\bGamma)$, the cross-sectional bootstrap is asymptotically valid. In sharp contrast,    in the model of
\cite{andrews2000inconsistency} and \cite{ketz2017testing}, $\Var(\bgamma_m|\bX)$ is explicitly modeled as an unknown parameter and appears in the loss function. Therefore, the discontinuity in the 
  asymptotic distribution explicitly arises from the presence of the boundary parameter in their loss functions, but is not 
    due to the issue of interplay among multiple terms in the asymptotic expansion.  Another reason why the cross-sectional bootstrap is valid in our context is that, as we illustrated in the above, the effect of estimating $\bgamma_m$ can be avoided by resampling the cross-sectional units, and the bootstrap variance directly estimates the asymptotic variance of term $(\bb)$. 
In contrast, the usual ``plug-in" method for the estimated asymptotic variance  is not uniformly valid  because the estimation error for   $\Var(\bgamma_m|\bX)$ dominates the  estimand.\footnote{A possible alternative approach is to employ the thresholding: estimate $\bV_{\gamma  }$ using $\widehat \bV_{\gamma}1_{\{\|\widehat \bV_{\gamma}\| <c_T \log T\}}$ for some   sequence $c_T\asymp \min\{T, \sqrt{p}\}^{-1}$, so that $c_T\log T$ ``just dominates"  $\|\bV_{\gamma} - \widehat\bV_{\gamma}\|$.     The similar approach has been employed to deal with the distribution discontinuity in the context of random coefficient models, and moment inequalities (e.g., \cite{andrews2010inference}). But in the current context, it has a few drawbacks. One is that it is hard to cover the entire space of all possible sequences for the eigenvalues of $\bV_{\gamma}$. It also leaves a question of choosing the constant in $c_T$. So we do not pursue it in this paper.}



The discussions in this section are based on a very simple setting, assuming that: (1) the model is discrete-time; (2)   the betas are time-invariant; (3) factors are directly observable; (4) the parameter of interest is finite dimensional;  (5) there are no drifts.  
In  Section \ref{sec: model} we shall formally explore this idea in a continuous-time conditional factor model with drifts using high-frequency data, and separately consider observed and estimated factors. We also 
extend the regression model to  more general high-dimensional  moment condition based models in Section \ref{s:gmm}.

 \section{The Continuous-Time   Factor Model with Characteristics}\label{sec: model}

\subsection{The model}

Consider a large panel of $p$ time series $\bY_t = (Y_{1t}, \cdots, Y_{pt})$, where $p$ is large. In financial asset pricing applications, $\bY_t  $ can be the vector of log-prices of $p$ stocks at time $t$. We assume $\bY=\{\bY_t\}_{t\geq 0}$ is a multivariate It\^{o} semimartingale on a filtered probability space $(\Omega, \calF, \{\calF_t\}_{t\geq0}, \bbP)$. 
For simplicity, we begin with a model without jumps, as we are interested in the continuous components of log-prices and factors. The jump-robust estimators are given in Section \ref{subsec:Jump-Robust}, where we employ  a standard procedure to truncate jumps out.\footnote{In addition, we assume there is no micro-structure noises. In  empirical studies we use data of five-min frequency.  In the presence of micro-structure noises, other solutions include sub-sampling (\cite{Zhang&MA:2005}), realized kernel (\cite{BN-HLN:2008}) and pre-averaging (\cite{Jacod&LMPV:2009}). Our main results  remain valid when using those more complicated noise-robust estimators.  }

We assume the following (continuous) factor structure:
\begin{gather}
\begin{split}  
    \bY_t =\bY_0 &+ \int_0^t \balpha_s ds + \int_0^t \bbeta_s d\bF_s + \bU_t
\end{split}
\end{gather}
where $\bY_0$ is the starting value of the process $\bY$ at time 0, the drift process $\balpha = \{\balpha_s\}_{s\geq0}$ is an optional $\RR^p$-valued process, the factor loading process $\bbeta=\{\bbeta_t\}_{t\geq0}$ is an optional $p\times K$ matrix process. The $K$ dimensional continuous factor process $\bF_t$ and the idiosyncratic continuous risk $\bU_t$ 
can be represented as 
\begin{align} \label{eq:FU}
\begin{split}
	\bF_t &=\int_0^t \balpha_s^F ds + \int_0^t \bsigma_s^F d\bW^F_s, \cr
	\bU_t &= \int_0^t \bsigma_s^U d\bW_s^U,
\end{split}
\end{align}
 where $\bW^U$ and $\bW^F$ are two multi-dimensional Brownian motions and are orthogonal  to each other (that is, their quadratic covariation is zero), and $\balpha^F = \{\balpha_s^F\}_{s\geq0}$ is the drift process of  the factors.  At any time point $t$, we write $\bbeta_t=(\bbeta_{1t},...,\bbeta_{pt})'$ and in general, each $\bbeta_{lt}$ ($l=1,\cdots,p$) is a $ K\times 1$ vector of adapted stochastic processes. In the literature, this beta is referred to  as the continuous beta (\cite{BLT:2016}). 

In addition, for each firm $l\leq p$, we observe  a set of  (possibly) time-varying characteristics:
\begin{align*}
	 \bX_{lt}=(\bx_{l,t}', \bx_l', \bx_{t}')',\quad l=1,\cdots,p.
\end{align*}
We allow the characteristics $\bX_{lt}$ to consist of (1) common  time-varying characteristics $\bx_{t}$    (such as term and default spread and macroeconomic variables); (2) individual specific characteristics $\bx_{l}$ that are time-invariant over the sampling period $[0,T]$ (such as  size and value which change annually); and (3) characteristics $\bx_{l,t}$ that are both  time-varying and individual specific. Here we present $(\bx_{l,t}', \bx_l', \bx_{t}')'$ with a bit  abuse of notation.

In this paper, we consider the following decomposition of the  continuous betas:
\begin{align} \label{eq:Beta}
   \bbeta_{lt} = \bg_{t}(\bX_{lt}) + \bgamma_{lt}, \quad l=1,\cdots,p.
\end{align}
The overall effect of characteristics on the factor loadings is represented by $\bg_{lt}:=\bg_{t}(\bX_{lt}) $, and is called ``characteristic beta".
Here $\bg_{t}(\cdot) $ is   a nonparametric function of macroeconomic and firm  variables,  possessing  less volatile  and picks  up long run beta fluctuations.
On the other hand, $\bgamma_{lt}$   represents the remaining time-varying individual factor risks after conditioning on the observed characteristics,  and captures high frequency movements in betas.  
The two components capture different aspects of  beta dynamics.
For the identification purpose, we assume $\E(\bgamma_{lt}| \bX_{t})=0$ for $\bX_t=\{\bX_{lt}: l\leq p\}$, which well separates the characteristic effects from the remaining effects.  The goal is to make uniform inference about $\bg_{lt}$  over a broad  DGP class, which admits various 
 cross-sectional variations in $(\bgamma_{lt}$, $\bg_{lt})$ and  dynamics.

We separately      study  two cases: known and unknown factor cases. In the ``known factor case", the high-frequency return data
of the common factors are observable, as in the case of \cite{ait2014idiosyncratic}  who constructed Fama-French factors using high-frequency returns. On the other hand, the ``unknown factor case" refers to situations in which we do not observe the high-frequency factors,  but can   estimate  them  from a large continuous-time panel  (up to a locally time-invariant rotation matrix).  We shall  show that the effect of estimating  $\bF_t$  leads to an asymptotic bias that needs to be corrected.

\subsection{Discussion of the Condition $\E(\bgamma_{lt}|\bX_t)=0$}
  
The condition  $\E(\bgamma_{lt}|\bX_t)=0$ serves as a central condition to achieve the identification of the characteristic effects, under which both components in the beta decomposition are well separated.       
  We now discuss the plausibility of this condition and possible approaches to relaxing it.  In the presence of omitted characteristics, say $\widetilde\bgamma_{lt}$, its ``explanable component" $\mathbb E(\widetilde\bgamma_{lt}|\bX_t)$ is ``absorbed"  in $\bg_{lt}$, and $\bgamma_{lt}$ only contains the orthogonal component $\widetilde\bgamma_{lt}- \mathbb E(\widetilde\bgamma_{lt}|\bX_t)$.



In the absence of  this condition, identification is lost, and we need further exogenous variables to identify the effect of characteristics.   Consider the ideal case  that $\bbeta_{lt}$ were   known.  Then in (\ref{eq:Beta}), $\bX_{lt}$ is endogenous.  To identify $\bg_t(\cdot)$, consider an instrumental variable approach: we need to find an exogenous instrumental variable $\bw_{lt}$ so that $\E(\bgamma_{lt}|\bw_{lt})=0.$ Define the operator:
$$
\mathcal T: \bg\to \E(\bg(\bX_{lt})|\bw_{lt}).
$$
We then have $\mathcal T(\bg_t)=\E(\bbeta_{lt}|\bw_{lt})$. The identification of $\bg_t$ depends on the invertibility of $\mathcal T$, and holds if and only if the conditional distribution of $\bX_{lt}|\bw_{lt}$ is complete, which is an untestable condition (see, e.g., \cite{newey2003instrumental}).  Suppose $\mathcal T$ is indeed invertible, it is well known that  estimating $\bg_t$ becomes an ill-posed inverse problem, and regularizations are needed, with possibly  a very slow rate of convergence.  We refer to the   literature for related estimation and identification issues: \cite{hall2005nonparametric, darolles2011nonparametric,  chen2012estimation}, etc. Therefore, while relaxing the condition $\E(\bgamma_{lt}|\bX_t)=0$ is  possible using the  nonparametric instrumental variable   approach, it requires a very different argument for the  identification and estimation. We do not pursue it in this paper. 

\subsection{Estimation}\label{s:estimate}

  Suppose there are in total $n$  sample intervals  on the interval $[0,T]$, with equal interval length, being   $\Delta_n$. 
 For any $t\in[0,T)$, define $I_t^n=\{ \lfloor t/\Delta_n \rfloor +1,\cdots,  \lfloor t/\Delta_n \rfloor + k_n \}$, where $ \lfloor \cdot \rfloor$ is the floor (greatest integer) function, and $k_n$ is the number of high frequency observations within the window $I_t^n$.      
 Ignoring the jumps,  fix any $t$, and for all  $i\in I_t^n$, by the Burkholder-Davis-Grundy inequality (cf. Chapter 2 of \cite{Jacod&protter:2011}), we have the following approximation for $\Delta_i^n \bY:= \bY_{i\Delta_n} - \bY_{(i-1)\Delta_n}$  (which is a $p\times 1$ vector):
\begin{align}\label{e:discretemodel}
\begin{split}
\Delta_i^n\bY &= \int_{(i-1)\Delta_n}^{i\Delta_n} \Big( \balpha_s ds + \bbeta_s d\bF_s + d\bU_s \Big)  \\
&= \balpha_{t}\Delta_n+\bbeta_{t}\Delta_i^n\bF + \Delta_i^n\bU + o_P(\Delta_n\sqrt{k_n}).
\end{split}
\end{align}
where {$o_P(\Delta_n\sqrt{k_n})$    } holds for each fixed element of $\Delta_i^n\bY$, and is uniform in $i\in I_t^n$.  Let $\bG_t$ be the $p\times K$ matrix of $\{ \bg_{lt} \}_{l=1}^p$ and $\bGamma_t$ be the $p\times K$ matrix of $\{\bgamma_{lt}\}_{l=1}^p$.   We shall then use all observations on $I_t^n$ to estimate $\bG_{t}$. 
Let $\dim(\bbeta_{lt})= K$,   and $\dim(\bX_{lt})=K_x$. Throughout the paper, we shall assume $p,  k_n\to\infty$, $\Delta_n\to0$, while $K,    K_x$  are fixed constants.

To define the estimators of $\bGamma_t$ and $\bG_t$, we introduce the following notation.   Let $\bphi_{lt}=(\phi_1(\bX_{lt}),...,\phi_J(\bX_{lt}))'$ be a $J\times 1$ vector of sieve basis functions of $\bX_{lt}$, which can be taken as, e.g., Fourier basis, B-splines, and wavelets. Let $\bPhi_t=(\bphi_{1t },...,\bphi_{pt})'$ be the  $p\times J$ basis matrix, and define the projection matrix:
 $$
\bP_t=\bPhi_t(\bPhi_t'\bPhi_t)^{-1}\bPhi_t',\quad p\times p.
$$
 We subsequently discuss the estimation procedures for the known and unknown factor cases.

\subsubsection{The Known Factor Case}

 In the known factor case, we also observe $\{\Delta_i^n\bF\}_{i\in I_t^n}$ in each interval. 
We use the following two-step estimation:

\textit{Step 1.} 
Run time-series regression:
\begin{equation}\label{zestimation}
\widehat\bbeta_{t}:= \sum_{i\in I_t^n}  \Delta_i^n\bY \, \Delta_i^n\bF' \left(\sum_{i\in I_t^n} \Delta_i^n\bF \, \Delta_i^n\bF' \right)^{-1}
\end{equation}
Write $\widehat\bbeta_{t}=(\widehat\bbeta_{1t},...,\widehat\bbeta_{pt})'$ to be the $p\times K$ matrix. 
 
 \textit{Step 2.}   Run cross-sectional regression:
  \begin{eqnarray*}
 \widehat\bG_t&=&(\widehat\bg_{1t},...,\widehat\bg_{pt})':=\bP_t\widehat\bbeta_{t},\cr
  \widehat\bGamma_t&=&(\widehat\bgamma_{1t},...,\widehat\bgamma_{pt})':=(\bI_p-\bP_t)\widehat\bbeta_{t}  .
\end{eqnarray*}

Putting together, the estimators can be expressed as:
\begin{align}\label{e3.2}
\begin{split}
 \widehat \bG_t &= \sum_{i\in I_t^n} \bP_{t} \, \Delta_i^n\bY \, \Delta_i^n\bF' \left(\sum_{i\in I_t^n} \Delta_i^n\bF \, \Delta_i^n\bF' \right)^{-1},\cr
  \widehat \bGamma_t &= \sum_{i\in I_t^n}(\bI_p-\bP_{t}) \Delta_i^n\bY \, \Delta_i^n\bF' \left(\sum_{i\in I_t^n} \Delta_i^n\bF \, \Delta_i^n\bF' \right)^{-1}.
\end{split}
\end{align}

\subsubsection{The Unknown Factor Case}

When factors are unknown, we first estimate the latent factors and  use them in place of $\Delta_i^n\bF$ in  (\ref{e3.2}). In the continuous-time factor model literature such as  \cite{ait2017using} and \cite{Pelger:2016},  these factors are estimated using  the regular principal component (PCA) method, extended from  \cite{connor1986performance}. But different from these works, we employ the PCA on the ``projected  returns", a method that was proposed by \cite{fan2016projected} in the discrete unconditional factor model. 
Here we extend the  method of this procedure to the continuous-time conditional factor model, and study its asymptotic effect for inference about the characteristic betas. 

We use the  simplified notation $\bP_{i-1}:=\bP_{(i-1)\Delta_n}$, $\bG_{i-1}:=\bG_{(i-1)\Delta_n}$, and $\bGamma_{i-1}:=\bGamma_{(i-1)\Delta_n}$. 
On each local window $I_t^n  $, we can define the following $p\times k_n$ matrix:
\begin{align*}
(\bP\Delta^n\bY)_t = (\bP_{i-1}\Delta_i^n\bY: i\in I_t^n) 
 \end{align*}
 Define the estimated factors, a $k_n\times K$ matrix
$$
\widehat{\Delta^n\bF}=(\widehat{\Delta_i^n\bF}: i\in I_t^n)', 
\quad k_n\times K,
$$
whose columns equal  $\sqrt{k_n\Delta_n}$ times the eigenvectors of the $k_n\times k_n$ matrix $\frac{1}{pk_n\Delta_n}(\bP\Delta^n\bY)_t'(\bP\Delta^n\bY)_t$ 
 corresponding to the first $K$ eigenvalues. 
We then use  estimated factors in place of $\{\Delta_i^n\bF\}_{i\in I_t^n}$ in (\ref{e3.2}):
\begin{eqnarray}\label{e3.5}
 \widehat \bG_t^{\latent}&=&\bP_{t}\frac{1}{k_n\Delta_n} \sum_{i\in I_t^n} (\Delta_i^n\bY) \widehat{\Delta_i^n\bF}',\cr
  \widehat \bGamma_t^{\latent}&=&(\bI_p-\bP_{t}) \frac{1}{k_n\Delta_n}  \sum_{i\in I_t^n}(\Delta_i^n\bY) \widehat{\Delta_i^n\bF}'
 \end{eqnarray}
and note that $\frac{1}{k_n\Delta_n}\sum_{i\in I_t^n} \widehat{\Delta_i^n\bF} \, \widehat{\Delta_i^n\bF}' =\bI_K$.
The $l$-th components  $\widehat\bg_{lt}^{\latent} $ and $\widehat\bgamma_{lt}^{\latent} $, respectively  estimate the characteristic   and idiosyncratic betas for the $l$-th individual.
The superscript ``latent" indicates that the estimators are defined for the case of latent factors.

 We now give an intuitive explanation on the estimated factors.  Note that $\bP_{i-1}$ is the cross-sectional projection matrix onto the  space expanded by  the sieve transformations of the characteristics at time $(i-1)\Delta_n$. 
   Apply the projection to the discretized model:
\begin{align*}
\bP_{i-1} \Delta_i^n \bY =\,& \bG_{i-1} \Delta_i^n\bF + \underbrace{\bP_{i-1} \bpsi_{i-1} \Delta_n}_{\text{higher-order term}} +\underbrace{\bP_{i-1} \bGamma_{i-1} \Delta_i^n\bF + \bP_{i-1} \Delta_i^n \bU}_{\text{projection errors}} \cr
 	& + \underbrace{(\bP_{i-1} \bG_{i-1} - \bG_{i-1}) \Delta_i^n\bF}_{\text{sieve approximation errors}}, 
 \end{align*}
 where $\bpsi_{i-1}$ denotes the higher-order drift and diffusion terms.
By the identification conditions  $\EE(\bGamma_t|\bX_t)=0$ and that $\EE( \bU_{t+s} - \bU_t | \bX_t)=0$,  the two components of the ``projection errors" are projected off, whose rate of decay (after standardized by $\Delta_n^{-1/2}$) is of $O_P(p^{-1/2})$. 
 Ignoring the    higher-order term, we have
\begin{equation}\label{e:ols}
\bP_{i-1} \Delta_i^n\bY \approx \bG_{i-1} \Delta_i^n\bF, 
\end{equation}
which is nearly ``idiosyncratic-free", and leads to 
$$
(\bP\Delta^n\bY)_t'(\bP\Delta^n\bY)_t\approx\Delta^n\bF \bG_t'\bG_t\Delta^n\bF'.
$$
Therefore  the columns of $\Delta^n\bF$ are approximately the eigenvectors of the ``idiosyncratic-free" matrix $(\bP\Delta^n\bY)_t'(\bP\Delta^n\bY)_t$, up to a rotation. 
Hence we can estimate them by applying  PCA on   $(\bP\Delta^n\bY)_t'(\bP\Delta^n\bY)_t$.  

Furthermore,    (\ref{e:ols}) also yields:
\begin{align}\label{e:gammaeffect}
(\bI_p-\bP_{i-1})\Delta_i^n\bY  \approx \bGamma_{i-1} \Delta_i^n\bF + \Delta_i^n\bU  . 
\end{align}
It  shows that $\bGamma_{i-1}$  represents the  loading on the risk factors of the remaining components of returns, after the characteristic effect is conditioned.

\subsubsection{Jump-robust estimators } \label{subsec:Jump-Robust}
In the general case with jumps, we employ the truncation method to remove those jumps. For notation simplicity, we omit the details and simply assume the jumps are of finite variation.  In the known factor case, we replace each $\Delta_i^n \bY$ and $\Delta_i^n \bF$ (previously assumed to be continuous) with their truncated versions:
\begin{align*}
 \widehat \bG_t &= \sum_{i\in I_t^n} \bP_{i-1} \, \Delta_i^n\bY_{\psi_n^Y} \, \Delta_i^n\bF'_{\psi_n^F} \left(\sum_{i\in I_t^n} \Delta_i^n\bF_{\psi_n^F} \, \Delta_i^n\bF'_{\psi_n^F} \right)^{-1},\cr
  \widehat \bGamma_t &= \sum_{i\in I_t^n}(\bI_N-\bP_{i-1}) \Delta_i^n\bY_{\psi_n^Y} \, \Delta_i^n\bF'_{\psi_n^F} \left(\sum_{i\in I_t^n} \Delta_i^n\bF_{\psi_n^F} \, \Delta_i^n\bF'_{\psi_n^F} \right)^{-1},
\end{align*}
where $\Delta_i^n Z_{\psi_n^Z}:=\Delta_i^n Z_l \, 1_{\{ \| \Delta_i^n Z_l \| \leq \psi_n^{Z_l} \}}$ denotes the usual truncated process for the process $\Delta_i^nZ$, with some random sequence $\psi_n^Z$ that depends on certain property of $Z$ and converges in probability to zero as $\Delta_n\rightarrow0$ (e.g., \cite{Mancini:2001}).\footnote{The common practice is the set $\psi_n^{Z_l} = \alpha_l \Delta_n^\varpi$, where $\varpi\in(0,1/2)$, $\alpha_l=C (\frac{1}{t} \text{IV}(Z_l)_t)^{1/2}$ with $C=3,4$ or $5$ and $\text{IV}(Z_l)_t$ is the integrated volatility of $Z_l$ over $[0,t]$.}
 In the unknown factor case, we only need to replace each $\Delta_i^n \bY$ with its corresponding truncated versions.

\subsection{Assumptions}
We now present the technical assumptions  for the asymptotic properties of the estimated characteristic effect $\widehat\bg_{lt}$ for a fixed $l\leq p$. This subsection presents   the required conditions to prove the limiting distribution. In particular, we  allow  the idiosyncratic components $\{\bU_t, \bGamma_t\} $ to be cross-sectionally weakly dependent.  In Section \ref{sec:bootstrap:factor}, we shall present the required conditions for the cross-sectional bootstrap, where we assume them to be cross-sectionally independent.

 We assume that   the following conditions hold uniformly over a class of DPG's: $\mathbb P\in\mathcal P$.    We apply the standard assumptions to define the stochastic processes as follows (e.g.,   \cite{Protter:2005}).

\begin{ass}[Data Generating Process] \label{a:dgp}
(i) The process $\bY$ is an It\^{o} semimartingale, whose continuous component is given by \eqref{eq:Y}, and  the continuous component of $\bF$ and $\bU$ are given by \eqref{eq:FU}.  The jump components of $\bY$ and $\bF$ are of finite variation. 

(ii) Write  $\btheta_{lt}:= ( \bX_{lt}, \bgamma_{lt}, \alpha_{lt}, \sigma^{U}_{lt}, \bsigma^{F}_t )$. For each $l\leq p$, $\btheta_{lt}$ is a multivariate continuous and locally bounded (uniformly in $l\leq p$) It\^{o} semimartingale with the form:   
\begin{align*}
	\btheta_{lt} =\, & \btheta_{l0} + \int_0^t \tilde{\balpha}_{ls} ds + \int_0^t \tilde{\bsigma}_{ls} d\bW_s^l  .
\end{align*}
Here $ \{\widetilde\balpha_{ls}\}_{s\geq0}$ and $ \{\widetilde{\bsigma}_{ls}\}_{s\geq0}$   are optional processes and locally bounded uniformly in $l\leq p$. We allow $\bW^l_t$ to be correlated with $(\bW_t^F, \bW^U_t)$ defined in (\ref{eq:FU}).  



(iii) $\E(\bgamma_{lt}|\bX_{t})=0$ for all $t\in[0, T), l\leq p$.

(iv) The quadratic covariation $[\bF, \bU]_t=0$ for all $t$.  
\end{ass}




\begin{ass}[Smoothness with respect to time]\label{a3.2:smooth:g} 
There are absolute constants  $C, \eta>0$,

(i) Write  $ \widetilde \bg(t, \bx):=\bg_{t}(\bx)$. Then $\widetilde\bg $ and $\phi_j$ are differentiable, satisfying $$
\sup_{t\in [0, T], \bx\in\mathcal X}\left|\frac{\partial\widetilde \bg(t, \bx)}{\partial t}  \right| +\sup_{t\in [0, T], \bx\in\mathcal X}\left|\frac{\partial\widetilde \bg(t, \bx)}{\partial \bx}  \right|
+\max_{l\leq N, j\leq J, \bx\in\mathcal X}  \left|\frac{\partial\phi_j(\bx)}{\partial \bx}  \right|
<C,
$$
where $\mathcal X$ is the domain of $\{\bX_{lt}\}_{l\leq p, t\in[0, T]}$. 

 (ii)  $\max_{t\in[0, T]}\|\bG_t-\bP_t\bG_t\|_\infty\leq CJ^{-\eta}$ almost surely for   $\eta>0$ so that $J^{-\eta} \sqrt{pk_n}=o(1)$.

\end{ass}
The above  assumption ensures that  $\bg_{t}(\bX_{lt})$ and $\bphi_{lt}$ are smooth transformations of $\bX_{lt}$. In particular, condition (i) is regarding the smoothness with respect to time,  so $\bg_{t}(\bX_{lt})$ and $\bphi_{lt}$  are also semimartingales; Condition (ii) is regarding the smoothness with respect to cross-sections.   It holds if $\max_{t, \bx}\|\bg_t(\bx)-\sum_{j=1}^J\bb_j\phi_j(\bx)\| \leq CJ^{-\eta}$
for some sieve coefficients $\{\bb_j: j\leq J\}$.

We now describe the asymptotic variance of $\widehat \bg_{lt}$, and introduce further notation. Let $\bc_{FF,t}$ ($K\times K$) and $\bc_{uu,t}$ ($p\times p$) be the instantaneous quadratic variation processes of $\bF=\{\bF_t\}_{t\geq0}$ and $\bU=\{\bU_t\}_{t\geq0}$. Let $U_{l,t}$ denote the $l$-th component of $\bU_t$, where $l\leq p$.  Also,  let    
$$
h_{t,lm} =\bphi_{lt}'(\frac{1}{p}\bPhi_t'\bPhi_t)^{-1}\bphi_{mt},\quad l,m\leq p.
$$
 The asymptotic variance depends on the following matrices.
  \begin{eqnarray*}
\bV_{u, t} &=&  \lim_{s\to0}\bc_{FF,t}^{-1}\Var\left( \frac{1}{ s\sqrt{p}}\sum_{m=1}^p h_{t,ml}(\bF_{t+s}-\bF_t)(U_{m,t+s}-U_{m,t})\bigg{|}\mathcal F_t\right)  \bc_{FF,t}^{-1}\cr
\bV_{\gamma, t} &=& \Var\left(\frac{1}{\sqrt{p}}\sum_{m=1}^ph_{t,ml}\bgamma_{mt}\bigg{|}\bX_t\right).
\end{eqnarray*}

\begin{ass}[Moment Bounds]\label{a:sieve}
There are  absolute constants $c, C>0 $, so that 
 \\
 (i) $\max_{l\leq p, t\in[0, T]}\E\|\bg_{t}(\bX_{lt})\|^4<C$, $\E\|\bgamma_{lt}\|^4<C$. \\
(ii) $\max_{t\in[0, T], l,m\leq p}\E h_{t,lm} ^4\leq C.$\\
(iii) $c<\min_{t\leq T}\lambda_{\min}(\frac{1}{p}\bPhi_t'\bPhi_t)\leq \max_{t\leq T}\lambda_{\max}(\frac{1}{p}\bPhi_t'\bPhi_t)<C $  almost surely.
\end{ass}


\begin{ass}[Cross-sectional Weak Dependence] \label{a:errorcov}  

  There are  absolute constants $c, C>0$  so that almost surely,
  
  (i)     $\max_{m\leq p}\sup_{t\in[0, T]}\Var(\bgamma_{mt}|\bX_t)<C$, and
$$
c< \inf_{t\in[0,T]}\lambda_{\min}(\bV_{u,t})\leq  \sup_{t\in[0,T]}\lambda_{\max}(\bV_{u,t})<  C,\qquad 
 \lambda_{\max}(\bV_{\gamma,t}) \leq  C\lambda_{\min}(\bV_{\gamma,t}).$$

(ii)  If $\lambda_{\min}(\bV_{\gamma, t})\neq0$, then
$$
\bV_{\gamma, t}^{-1/2}\frac{1}{\sqrt{p}} \sum_{m=1}^p\bgamma_{mt}h_{t,ml}\ConStable N(0, \bI_K).
$$

(iii) Uniformly in $t\in[0,T]$, almost surely, $ \|\bc_{uu,t}\|<C$, $ \lambda_{\min}(\bc_{FF,t})>c $  and    
$$
 \max_{s\in[0,\Delta_n]}\mathbb E\left( \left \|\frac{1}{s\sqrt{p}}\sum_{m=1}^p h_{t,ml}  (\bF_{t+s}-\bF_t)(U_{m,t+s}-U_{m,t})\right\|^4 \bigg{|}\mathcal F_{t}\right)<C.
$$

\end{ass}

The asymptotic distribution of $\widehat\bg_{lt}$ is jointly determined by two  uncorrelated components:  
\begin{equation}\label{e3.12} 
  \widehat\bg_{lt} -    \bg_{lt}    =
 \bc_{FF,t}^{-1} \frac{1}{k_n\Delta_n }\sum_{i\in I_t^n} \Delta_i^n\bF \Delta_i^n\bU'\bP_{t,l}
+ \bGamma_{t}' \bP_{t,l}   +  o_P((k_np)^{-1/2}) ,
\end{equation} where $\bP_{t,l} $ denotes the $l$-th column of $\bP_t$.
  Assumptions  \ref{a:errorcov} condition (ii)  requires that  $\bgamma_{lt}$ be cross-sectionally weakly dependent, and $ \bGamma_{t}' \bP_{t,l} $ admits a cross-sectional CLT. In addition, condition (iii) requires $\{U_{mt}: m\leq p\}$ be cross-sectionally weakly dependent. These conditions ensure the asymptotic normality of (\ref{e3.12}). 

\begin{ass}[For  estimated  factors]\label{a:latentfactor} Define $\bSigma_{G, t}= \frac{1}{p}\bG_t'\bG_t$. Almost surely:

(i) $c<\inf_{t\leq T}\lambda_{\min}(  \bSigma_{G, t})\leq \sup_{t\leq T}\lambda_{\max}(\bSigma_{G, t})<C $ for  absolute constants $c, C>0$.

(ii) The eigenvalues of $\bSigma_{G, t}^{1/2} \, \bc_{FF,t} \, \bSigma_{G, t}^{1/2}$ are distinct: there is an  absolute constant   $c >0$, so that the ordered  eigenvalues    $\lambda_1<\lambda_2...<\lambda_K$ of $\bSigma_{G, t}^{1/2} \, \bc_{FF,t} \, \bSigma_{G, t}^{1/2}$   satisfy:
$$
\lambda_{1}>c,\quad \lambda_{j+1} -\lambda_j>c,\quad j=1,..., K-1.
$$
\end{ass}

Assumption \ref{a:latentfactor} is similar to the \textit{pervasive condition} in the approximate factor model's literature, which identifies the latent factors (up to a rotation). In particular, condition (i) requires  that the betas should nontrivially depend on  $\bX_{t}$.

\subsection{Asymptotic Distributions and Uniform Bias Correction}
We first present the estimated  $\bg_{lt}$ for a fixed $(l,t)$ when factors are observable.  

\begin{thm}[known factor case]\label{t:g} Suppose $J, p, k_n\to\infty$ and $\Delta_n\to0$ satisfy: $J^2=O(p)$,   $k_npJ^{-2\eta} +pk_n^2\Delta_n =o(1)$. 
 Under Assumptions \ref{a:dgp}-\ref{a:errorcov},   for a fixed $(l,t)$,
$$
\left(\frac{1}{k_np}\bV_{u,t}+\frac{1}{p} \bV_{\gamma, t}\right)^{-1/2}\big(\widehat \bg_{lt}-\bg_{lt} \big) \ConStable N(0,\bI_K).
$$
\end{thm}

When the factors are latent and estimated, $\widehat \bg_{lt}$ consistently estimates a rotated $\bg_{lt}$. Up to the rotation, the asymptotic variance is identical to that of the known factor case. However,  the effect of estimating the factors gives rise to a bias term. Let $\widehat \bV_t$ be a  $k_n\times k_n$ diagonal matrix consisting of the first $K$ eigenvalues of  $\frac{1}{pk_n\Delta_n}(\bP\Delta^n\bY)_t'(\bP\Delta^n\bY)_t$. Let
\begin{align*}
\bM_t &=\frac{1}{k_n\Delta_n\sqrt{p}} \sum_{i \in I_t}   \widehat\bV_t^{-1} \,  \widehat{\Delta_i^n\bF} \, \Delta_i^n \bF' \, \bbeta_{(i-1)\Delta_n}' \bP_{i-1} \cr
\BIAS_g&=\bM_t \frac{1}{k_n \sqrt{p}}\sum_{i\in I_t} \bP_{i-1} \, \bc_{uu,(i-1)\Delta_n} \, \bP_{i-1,l}.
\end{align*}
 
We have the following theorem.
\begin{thm}[unknown factor case]\label{t:glatent} In addition to the assumptions for   Theorem \ref{t:g}, assume Assumption \ref{a:latentfactor} and $k_nJ=o(p^2)$. Then   there is a $K\times K$ rotation matrix $\bUpsilon_{nt}$, 
$$
\bUpsilon_{nt}^{'-1/2}\left(\frac{1}{k_np} \bV_{u,t} +\frac{1}{p} \bV_{\gamma, t} \right)^{-1/2}\bUpsilon_{nt}^{-1/2}\left(\widehat \bg_{lt}^{\latent}-\bUpsilon_{nt}\bg_{lt}-\BIAS_g\right)\ConStable N(0,\bI_K).
$$
\end{thm}

 We make several remarks.

\begin{rem}\label{r5.1}
If $\|\bV_{\gamma,t}\|=o_P(k_n^{-1}) $, 
then the rate of convergence is $O_P((k_np)^{-1/2})$, and  
\begin{align*}
\bV_{u,t}^{-1/2}\sqrt{k_np} \left(\widehat \bg_{lt}-\bg_{lt}\right)\ConStable N(0,\bI_K).
\end{align*}  Intuitively, this occurs when the cross-sectional variation of $\bgamma_{lt}$ is weak. As a result, the observed characteristics capture most of the beta fluctuations in $\bbeta_t$, leading to a fast rate of convergence.
 On the other hand,  if  $\lambda_{\min}(\bV_{\gamma, t})\gg k_n^{-1}$, 
\begin{eqnarray*}
\bV_{\gamma,t}^{-1/2}\sqrt{p} \left(\widehat \bg_{lt}-\bg_{lt}\right)&\ConStable&N(0,\bI_K).
\end{eqnarray*}
In particular, the rate of convergence is  $O_P(p^{-1/2})$ if the eigenvalues of $\bV_{\gamma, t}$ are bounded away from zero, corresponding to the case of strong cross-sectional variations in $\bGamma_t.$
Intuitively, this  means when idiosyncratic betas  have  strong cross-sectional variations, time-series regression in step 1 is not informative to estimating $\bg_{lt}$, and the main statistical error arises from the step 2 cross-sectional regression.  This leads to a slower rate of convergence.  \end{rem}

\begin{rem}
As in the known factor case, the two sources of the randomness  $\frac{1}{k_np} \bV_{u,t} $ and $\frac{1}{p} \bV_{\gamma, t} $  jointly determine  the rate of convergence and the  limiting distribution of $\widehat\bg_{lt}^{\latent}$.   In addition, estimating the latent factors leads to a bias term, whose order is      $O_P(p^{-3/2})$.  If $\|\bV_{\gamma,t}\|=o_P(k_n^{-1}) $,
\begin{eqnarray*}
(\bUpsilon_{nt}\bV_{u,t}\bUpsilon_{nt}')^{-1/2}\sqrt{k_np} \left(\widehat \bg_{lt}^{\latent}-\bUpsilon_{nt}\bg_{lt}-\BIAS_d\right) \ConStable N(0,\bI_K).
\end{eqnarray*}
But if $\lambda_{\min}(\bV_{\gamma, t})\gg \max\{k_n^{-1}, p^{-2}\}$, then $\widehat \bg_{lt}^{\latent}$ is {asymptotically unbiased}:
\begin{eqnarray*}
 (\bUpsilon_{nt}\bV_{\gamma,t}\bUpsilon_{nt}')^{-1/2}\sqrt{p} \left(\widehat \bg_{lt}^{\latent}-\bUpsilon_{nt}\bg_{lt}\right)&\ConStable& N(0,\bI_K).
\end{eqnarray*}
Therefore when the signal from $\bgamma_{lt}$ is sufficiently strong, the rate of convergence slows down, and dominates the bias arising from the effect of estimating factors. However, as the magnitude of the eigenvalues of $\bV_{\gamma, t}$ is unknown and may change over time, we nevertheless need a  bias-correction procedure.
\end{rem}

Generally, while $\|\bV_{\gamma,t}\|=o_P(k_n^{-1}) $ and $\lambda_{\min}(\bV_{\gamma, t})\gg k_n^{-1}$ are two special cases, we do not know the actual strength of the eigenvalues of $\bV_{\gamma,t}$ in practice. In fact, its eigenvalues can be any sequences in a large range, resulting in an unknown     rate of convergence for $ \left(\widehat \bg_{lt}-\bg_{lt}\right)$ between $O_P((k_np)^{-1/2})$ and $O_P(p^{-1/2})$. This calls for a need of uniform inference.  
We shall  rely on the cross-sectional bootstrap, as we formally present in the next subsection.

We now derive a bias-corrected
spot estimated $\bg_{lt}$ in the case of estimated factors. The bias correction  is valid uniformly over various signal strengths. 
Note that in $\BIAS_g$,   $\bM_t$ can be naturally estimated by $\widehat\bM_t=\frac{1}{ \sqrt{p}} \widehat\bV_t^{-1}  \widehat \bG_t'.$  The major challenge arises in  estimating the $p\times p$ quadratic variation $\bc_{uu, t}$, which is high-dimensional when $p$ is large.
We consider two  cases for the bias correction.

\textbf{CASE I:  cross-sectionally uncorrelated}

When $\{\Delta_i^n U_1,\cdots,\Delta_i^n U_p\}$ are  cross-sectionally uncorrelated, the $\calF_{i-1}$ conditional quadratic variation $\bc_{uu,(i-1)\Delta_n}$ is a diagonal matrix. Let $
\widehat{\Delta_{i}^n\bU}=\Delta_{i}^n\bY-(\widehat\bG_{i-1}+\widehat\bGamma_{i-1})\Delta_{i}^n\bF$.
Apply  \cite{white1980heteroskedasticity}'s  covariance estimator using the residuals:
$$
\widehat{\BIAS}_g=\widehat \bM_t \frac{1}{k_n\Delta_n\sqrt{p}} \sum_{i\in I_t^n} \bP_{i-1} \diag\{\widehat{\Delta_{i}^n\bU} \widehat{\Delta_{i}^n\bU}'\} \bP_{i-1,l}
$$

 \textbf{CASE II:  cross-sectionally weakly correlated  (sparse)}

In this case $\bc_{uu, (i-1)\Delta_n}$ is no longer diagonal. We shall assume it is a sparse covariance matrix, in  the sense that many of its off-diagonal entries are zero or nearly so. Then the thresholding estimator of \cite{POET} can be applied, yielding a nearly $\min\{k_n, p\}^{1/2}$- consistent sparse covariance estimator $\widehat{\bc}_{uu,t}$.
More specifically,  let $s_{dl}$ be the  $(d,l)$-th element of $\frac{1}{\Delta_nk_n}\sum_{i\in I_t^n}\widehat{\Delta_i^n\bU}\widehat{\Delta_i^n\bU}'$.
Let the $(d,l)$-th entry of the estimated covariance be:
$$
(\widehat \bc_{uu,t})_{dl}=\begin{cases}
s_{dd}, & \text{ if } d=l,\\
\text{th}(s_{dl})1_{\{|s_{dl}|>\varrho_{dl}\}} & \text{ if } d\neq l,
\end{cases}
$$
where $\text{th}(\cdot)$ is a thresholding function, whose typical choices are the hard-thresholding and soft-thresholding. 
 Here the threshold value $\varrho_{dl}= \bar C (s_{dd}s_{ll})^{1/2}\omega_{np}$ for some constant $\bar C>0$, with  $\omega_{np}=\sqrt{\frac{\log p}{k_n}} +\frac{J}{p}\max_{j,d}\frac{1}{J}\|\bphi(\bx_{jd})\|^2\sqrt{\log J}$.\footnote{Hard-threholding takes $\text{th}(s_{dl})=s_{dl}$, while soft-thresholding takes $\text{th}(s_{dl})=\sgn(s_{dl})( |s_{dl}|-\varrho_{dl} )$. We shall justify the choice of $\omega_{np}$ in Section \ref{sec:a.6}.  In addition, the choice of the constant $\bar C$ can be either guided using cross-validation, or simply a constant near one. For returns of S\&P 500, the rule of thumb choice $\bar C=0.5$ empirically works very well. We refer to \cite{POET} for more discussions on thresholding.}  Estimate the bias by:
$$
\widehat{\BIAS}_g=\widehat \bM_t\frac{1}{k_n\sqrt{p}}\sum_{i\in I_t^n} \bP_{i-1} \, \widehat{\bc}_{uu, t} \, \bP_{i-1,l}.
$$

 Formally,  we have  the following theorem for the bias-corrected estimator.
 \begin{thm}[Bias correction]\label{th:bias}
 Suppose  $Jk_n=o(p^2)$.  Consider CASE I and CASE II for estimated factors.  In particular, assume that for  $\max_{i\in I_t^n}\|\widehat{\bc}_{u,t} - \bc_{u,i} \|=o_P(\sqrt{\frac{p}{Jk_n}} )$\footnote{We shall verify this condition in Section \ref{sec:a.6} for sparse covariance estimators. } in CASE II.  Define the bias-corrected characteristic beta estimator
 $
\widetilde\bg_{lt}^{\latent} = \widehat \bg_{lt}^{\latent} - \widehat{\BIAS}_g.
 $
Under Assumptions \ref{a:dgp}-\ref{a:latentfactor}, we have
$$
\bUpsilon_{nt}^{'-1/2}\left(\frac{1}{k_np} \bV_{u,t} +\frac{1}{p}  \bV_{\gamma, t} \right)^{-1/2}\bUpsilon_{nt}^{-1/2}\left(\widetilde \bg_{lt}^{\latent} - \bUpsilon_{nt}\bg_{lt}\right) \ConStable N(0,\bI_K).
$$
 \end{thm}

\subsection{Uniform Confidence Intervals Using Cross-Sectional Bootstrap}\label{sec:bootstrap:factor}

\subsubsection{Independent cross-sectional bootstrap}

As we explained in Section \ref{sec:uniform}, the major difficulty is to handle    $\bGamma_t'\bP_{t,l}=\frac{1}{p}\sum_{m=1}^p \bgamma_{mt}h_{t,ml}$
in the asymptotic expansion of $\widehat\bg_{lt}-\bg_{lt}$, which contributes to the 
asymptotic variance  by  
$\frac{1}{p}\bV_{\gamma, t}$. If we use the standard ``plug-in"  method to estimate $\bV_{\gamma, t}$, we   would have to   introduce a $\gamma$-estimation  error 
  $$
 \frac{1}{p}\sum_{m=1}^p h_{t,ml}^2\left[ \widehat \bgamma_{mt}\widehat  \bgamma_{mt}'  -    \bgamma_{mt}  \bgamma_{mt}'  \right], 
$$
 which is not negligible  when $\bV_{\gamma, t}$ is near zero.  
Instead,  we propose  to use  cross-sectional bootstrap to  achieve uniform inference,  and let the bootstrap distribution ``mimic" cross-sectional variations.

 Let $\mathcal M:=\{m_1,..., m_p\}$ be a simple random sample with replacement from $\{1,..., p\}$, and we always fix $m_l= l$ when we are interested $\bg_{lt}$ for the $l$-th specific individual.  As we  do not need to mimic the time series variations, so for each sampled index $m\in \mathcal M$,  the entire time series $\{\Delta_i^nY_{m}, i\in I_t^n \}$  and $\{\bX_{m, i}, i\in I_t^n\}$  are  kept.  Therefore, we independently resample the  time series  and obtain the bootstrap data: $\{\Delta_i^nY_m, i\in I_t^n \}_{m\in \mathcal M}$ and $ \big\{\bX_{mi}: i\in I_t^n \big\}_{m\in\mathcal M}$. 
 In addition, we also keep the entire time series  $\{\Delta_i^n\bF \}$ in the case of known factors, and $\{\widehat{\Delta_i^n\bF} \}$ in the case of unknown factors.  The effect of estimating $\Delta_i^n\bF$ does not play a role in the cross-sectional variations. Hence we do not re-estimate the factors in each bootstrapped sample. 

  Let $\Delta_i^n\bY^*=(\Delta_i^nY_{m_1},...,\Delta_i^nY_{m_p})'$, $\bPhi_t^*=(\bphi_{m_1,t},...,\bphi_{m_p,t})'$ and $\bP_t^*=\bPhi_t^*(\bPhi_t^{*'}\bPhi_t^*)^{-1}\bPhi_t^{*'}$. Let $\bP_{t,l}^*$ be the $l$-th column of $\bP_t^*$. 
  Define
  \begin{eqnarray}\label{e:5.2}
 \widehat \bg_{lt}^*&=& \left(\sum_{i\in I_t^n} \Delta_i^n\bF  \Delta_i^n\bF'  \right)^{-1} \sum_{i\in I_t^n} \Delta_i^n\bF   \Delta_i^n\bY^{*'}  \bP_{t,l}^*,\quad \text{ when factors are known} \cr
 \widehat \bg_{lt}^{*\latent} &=&  \frac{1}{k_n\Delta_n} \sum_{i\in I_t^n} \widehat{\Delta_i^n\bF} \, \Delta_i^n\bY^{*'}  \bP_{t,l}^* ,  \quad \text{ when factors are estimated}
 \end{eqnarray}

When $\bg_{lt}$ is multidimensional, it is easier to present the confidence interval for a linear transformation   $\bv'\bg_{lt}$.
 The following algorithm summarizes the steps for computing the confidence intervals.
 
\begin{algo} Compute the confidence interval for   $\bv'\bg_{lt}$ (or $\bv'\bUpsilon_{nt}\bg_{lt}$ in the estimated factor case) as follows.
	
	Step 1.  Take a simple random sample $\{m_1,...,m_p\}$ with replacement from $\{1,...,p\}$.  Fix $m_l=l$.  Obtain $\Delta_i^n\bY^*=(\Delta_i^nY_{m_1},...,\Delta_i^nY_{m_p}: i\in I_t^n)'$ and $\bP_t^*$.
	
	Step 2.  Compute $ \widehat \bg_{lt}^{*}$, or $ \widehat \bg_{lt}^{*\latent}$ in the case of estimated factors, as in  (\ref{e:5.2}).
	
	Step 3. Repeat Step 1-2 for $B$ times  and obtain either $\{\widehat \bg_{lt}^{*b}\}_{b\leq B}$ or $\{\widehat \bg_{lt}^{*\latent, b}\}_{b\leq B}$, depending on whether factors are observable.
	For the  confidence level $1-\tau$, let  $q_{\tau}$ be the   $1-\tau$ bootstrap quantile of $\{  |  \bv'\widehat \bg^{*b}_{lt} -   \bv'\widehat \bg_{lt} |  \}_{b\leq B}$;  or $q^{\latent}_\tau$  as  the quantile of $\{  |  \bv'\widehat \bg^{*\latent, b}_{lt} -   \bv'\widehat \bg_{lt} ^{\latent}|  \}_{b\leq B}$.
	  
	  Step 4. Compute the confidence interval as:
	 \begin{eqnarray*}
	 	CI_{nt, \tau}&=&[\bv'\widehat \bg_{lt}  - q_\tau, \bv'\widehat \bg_{lt} +q_\tau], \cr
	 	(\text{or }  CI_{nt,\tau}^{\latent}&=& [\bv'\widehat \bg_{lt} ^{\latent} - \bv' \widehat{\BIAS}_g- q^{\latent}_\tau, \bv'\widehat \bg_{lt} ^{\latent} - \bv' \widehat{\BIAS}_g+ q^{\latent}_\tau]\quad ).
	 \end{eqnarray*}
\end{algo}

We need the following conditions for the bootstrap validity.

\begin{ass} \label{avfactor}  
 
Conditionally on $\bX_t$, $\{(\bgamma_{mt}, U_{m,t+s}-U_{mt}): m\leq p\}$ are cross-sectionally independent, 
for each $t\in[0,T]$ and $s\in[0,\Delta_n]$.

\end{ass}

\begin{ass}\label{a:gammabound} 
There are   absolute constants $C, \eta>0$, 

(i)  Almost surely,
\begin{eqnarray*}
  \frac{ \frac{1}{p}\sum_{m=1}^ph_{t,mm}^2h_{t,ll}^2\mathbb E(\|\bgamma_{mt}\|^4|\bX_t)}{\lambda^2_{\min}(\frac{1}{p}\sum_{m=1}^ph_{t,ml}^2\Var(\bgamma_{mt}|\bX_t))}<C.
\end{eqnarray*}

 If the denominator equals zero, then the above ratio is   defined as zero.

(ii) Almost surely in the bootstrap sampling space,  $\sup_{t\in[0,T]}\|\bG_t^*-\bP_t^*\bG^*_t\|_\infty\leq CJ^{-\eta}$, where $\bG_t^*=(\bg_{m_1, t},...,\bg_{m_p, t})'.$
 
\end{ass}
 
 Note that the bootstrap asymptotic variance for $\widehat \bg_{lt}^*$ is  analogously $ \frac{1}{k_np} \bV_{u, t} +\frac{1}{p} \widetilde{ \bV}_{\gamma, t}$, where $
 \widetilde{\bV}_{\gamma, t} =\frac{1}{p}\sum_{m=1}^ph_{i,ml}^2 \bgamma_{mt}  \bgamma_{mt}'$. The only approximation error for the $ \bV_{\gamma, t}$ part is the ``law of large numbers error":
$$
\widetilde{ \bV}_{\gamma, t}- \bV_{\gamma, t}=\underbrace{\frac{1}{p}\sum_{m=1}^ph_{t ,ml}^2 \left[\bgamma_{mt}  \bgamma_{mt}  ' -\Var(\bgamma_{mt}|\bX_t)\right]}_{\text{LLN error}}.
$$
Consequently, the $\gamma$-estimation error component is avoided.   This forms the foundation of the bootstrap asymptotic validity.   The moment  bound in Assumption \ref{a:gammabound} (i) on $\bGamma_t$  is used to ensure that the LLN-error is negligible regardless of the strength of $\bV_{\gamma,t}$.

 \begin{thm}[Uniformly valid confidence intervals]\label{t:conf:boot}
Let $\mathcal P$ be the collection of all data generating processes $\mathbb P$ for which   Assumptions \ref{a:dgp} -  \ref{a:gammabound}   hold.    Then for any fixed vector $\bv\in\mathbb R^{K}\backslash\{0\}$  such that $\|\bv\|>c>0$,  for each fixed $l\leq p, t\in[0, T]$, 
\begin{eqnarray*}
\text{known factor case: }&&\sup_{\mathbb P\in\mathcal P}\left| \mathbb P(\bv'\bg_{lt}\in CI_{nt,\tau})-( 1-\tau)\right|\to0\cr
\text{unknown factor case: }&& \sup_{\mathbb P\in\mathcal P}\left|\mathbb P(\bv'\bUpsilon_{nt}\bg_{lt}\in CI_{nt,\tau}^{\latent})-(1-\tau)\right|\to0.
\end{eqnarray*}
 \end{thm}

\subsubsection{Block cross-sectional bootstrap}

 Theorem \ref{t:conf:boot} requires the idiosyncratic components $\{\{\Delta_i^nU_m\}_{t\in [0, T]},\{\bgamma_{mt}\}_{t\in [0, T]}\}_{m\leq p}$  be cross-sectionally independent. 
We can relax this assumption to allow for cross-sectionally block-dependent idiosyncratic components when these blocks are known, and rely on block cross-sectional bootstraps. More specifically, suppose the cross-sectional index set has a non-overlapping partition $\{1,..., p\}=B_1\cup...\cup B_H$, with $H\to\infty$, and the cardinality of each ``block" $B_h$ is finite: $\max_{h\leq H}|B_h|_0=O(1).$ For a fixed $k\leq K$. We assume:
\\
(i) Individuals' block mememberships are known; (that is, $\{B_h\}_{h\leq H}$ are known)\footnote{In the more complicated case where blocks are unknown, one could first apply a block-thresholding method to estimate the cross-sectional covariance matrix of $\Delta_t^n\bU$, to consistently recover the block structures first. See, e.g., \cite{cai2012adaptive}.},
\\
(ii)  $\Cov(\Delta_t^nU_{l_1}, \Delta_t^nU_{l_2})=0$ and $\Cov(\gamma_{mt, k_1}, \gamma_{mt, k_2})=0$ for any $t\in[0,T]$ if the two individuals  $(l_1,l_2)$ belong to different blocks, for any $k_1, k_2\leq K$;   $\gamma_{mt,k}$ denotes the $k$-th element of $\bgamma_{mt}$. 

Therefore, conditionally on the factors, individuals are possibly correlated only within the same block. Empirically, the assumption of known   blocks of finite size can be supported by setting blocks as  industry sectors, which is motivated from the economic intuition that firms within similar industries are expected to have higher correlations conditioning on  the factors, e.g., \cite{ait2017using}. These blocks form a natural basis for the application of non-overlapping block bootstraps.  Suppose we are interested in the inference for $\bg_{lt}$ for a specific firm $l\leq p$, and it is known that $l\in B_{h_0}$ for a particular $h_0\leq H$. Set $l$ as the first element of $B_{h_0}$. We employ the block-bootstrap on the cross-sectional units, and can proceed with the following algorithm.


\begin{algo} Compute the confidence interval for $\bv'\bg_{lt}$ as follows. 

Step 1: Fix  $B_1^*= B_{h_0}$.   Take a simple random sample $\{B_2^*,..., B^*_H\}$  with replacement from $\{B_1,...,B_H\}$.

Step 2: For each sampled block $B_h^*$, all the individuals in the block are sampled associated with their entire time series. We obtain
\begin{eqnarray*}
\Delta_i^n\bY^*&=&(\Delta_i^nY_m:  m\in B_h^*, h\leq H),\quad i\in I_t^n\cr
\bPhi_t^*&=&(\bphi_{mt}:  m\in B_h^*, h\leq H).
\end{eqnarray*}
 and $\bP_t^*=\bPhi_t^*(\bPhi_t^{*'}\bPhi_t^*)^{-1}\bPhi_t^{*'}$. 
 
 Step 3: Define
 \begin{eqnarray*}
 	\widehat \bg_{lt}^*&=& \left(\sum_{i\in I_t^n} \Delta_i^n\bF  \Delta_i^n\bF'  \right)^{-1} \sum_{i\in I_t^n} \Delta_i^n\bF   \Delta_i^n\bY^{*'}  \bP_{t,l}^*\quad \text{ when factors are known}\cr
 	\widehat \bg_{lt}^{*\latent} &=&  \frac{1}{k_n\Delta_n} \sum_{i\in I_t^n} \widehat{\Delta_i^n\bF} \, \Delta_i^n\bY^{*'}  \bP_{t,l}^*   \quad \text{ when factors are estimated}.
 \end{eqnarray*}

Step 4:  Repeat Steps 1-3 for $B$ times,  and obtain either $\{\widehat \bg_{lt}^{*b}\}_{b\leq B}$ or $\{\widehat \bg_{lt}^{*\latent, b}\}_{b\leq B}$, depending on whether factors are observable.  Let  $q_{\tau}$ (or $q^{\latent}_\tau$) be the   $1-\tau$ bootstrap quantile of $\{  |  \bv'\widehat \bg^{*b}_{lt} -   \bv'\widehat \bg_{lt} |  \}_{b\leq B}$ (or  $\{  |  \bv'\widehat \bg^{*\latent, b}_{lt} -   \bv'\widehat \bg_{lt} ^{\latent}|  \}_{b\leq B}$).  Compute the confidence interval for $\bv'\bg_{lt}$ (or $\bv'\bUpsilon_{nt}\bg_{lt}$ in the estimated factor case) as:
\begin{eqnarray*}
	CI_{nt, \tau}&=&[\bv'\widehat \bg_{lt}  - q_\tau, \bv'\widehat \bg_{lt} +q_\tau], \cr
	(\text{or }  CI_{nt,\tau}^{\latent}&=& [\bv'\widehat \bg_{lt} ^{\latent} - \bv' \widehat{\BIAS}_g- q^{\latent}_\tau, \bv'\widehat \bg_{lt} ^{\latent} - \bv' \widehat{\BIAS}_g+ q^{\latent}_\tau]\quad ).
\end{eqnarray*}
\end{algo}
 
 The proof of the first-order bootstrap validity is very similar to that of Theorem \ref{t:conf:boot}, building on the  results of the validity of block-bootstrap \citep{andrews2004block, lahiri1999theoretical}. We omit the formal proof for technical simplicity.

 \subsection{Inference about the integrated characteristic-beta}

We can also estimate the long-run characteristic effect: $\int_0^T\bg_{lt}dt$ in the case of known factors.\footnote{Due to the rotation discrepancy, estimating the long-run effect with unknown factors subjects to the issue of time-varying rotations, and is a very challenging problem in the presence of time-varying betas, and we shall leave it for the  future research.}    Using the standard overlapping spot estimates (see, e.g. \cite{Jacod&Rosenbaum:2013}), we estimate it by
$$\widehat{\int_0^T\bg_{lt}dt}:=\sum_{t=1}^{[T/\Delta_n]-k_n}\widehat \bg_{lt}\Delta_n.$$
Asymptotically, it has the following decomposition:
\begin{eqnarray*}
 \widehat{\int_0^T\bg_{lt}dt}
- \int_0^T\bg_{lt}dt&=& \sum_{t=1}^{[T/\Delta_n]-k_n}\Delta_n  \bGamma_t'\bP_{t,l} 
+ \sum_{t=1}^{[T/\Delta_n]-k_n}\Delta_n\bc_{FF,t}^{-1}\frac{1}{k_n\Delta_n}\sum_{i\in I_t^n} \Delta_i^n\bF \Delta_i^n\bU'\bP_{t, l}  \cr
&&
+o_P\left(\sqrt{\frac{\Delta_n}{ p}}\right).
\end{eqnarray*}
As in the spot estimation case, the asymptotic expansion also admits two components: the effect of estimating  integrated betas  and the effect of cross-sectional estimation. 
The final limiting distribution is determined by the interplay of both terms. Due to the unknown signal strength of the idiosyncratic beta $ \sum_{t=1}^{[T/\Delta_n]-k_n}\Delta_n  \bGamma_t'\bP_{t,l}$, we still rely on the bootstrap to make uniform inference about $\bv'\int_0^T\bg_{lt}dt$ for any specific vector $\bv$ of interest.


 Denote by  $\widehat{\int_0^T\bg_{lt}dt}^{*b}=\sum_{t=1}^{[T/\Delta_n]-k_n}\widehat \bg_{lt}^{*b}\Delta_n$ as the bootstrap estimator in the $b$-th generated sample. Let
 $\widetilde q_{\tau}$  be the   $1-\tau$ bootstrap quantile of $\{  |  \bv'\widehat{\int_0^T\bg_{lt}dt}^{*b} - \bv'\widehat{\int_0^T\bg_{lt}dt}  |  \}_{b\leq B}$.  The   confidence interval for $\bv'\int_0^T\bg_{lt}dt$  is given by
\begin{eqnarray*}
\widetilde{CI_{n, \tau}}=\left[\bv'\widehat{\int_0^T\bg_{lt}dt}   - \widetilde q_\tau, \bv'\widehat{\int_0^T\bg_{lt}dt}   +  \widetilde q_\tau\right].
 \end{eqnarray*}

The following condition plays a similar role as that of Assumption \ref{a:gammabound}, but for estimating the integrated characteristic effects.
\begin{ass}\label{a:gammabound:longrun} There is an absolute constant $C>0$, almost surely, $$
  \frac{1}{p}\sum_{m=1}^p\frac{1}{[T/\Delta_n]}  \sum_{t=1}^{[T/\Delta_n]-k_n}  \mathbb E( \| \bgamma_{mt}\|^4h_{t,mm}^2h_{t,ll}^2|\bX_t)\leq C   
 \lambda_{\min}^{2}\left[ \frac{1}{ p} \sum_{m=1}^p   \Var( \frac{1}{[T/\Delta_n] }  \sum_{t=1}^{[T/\Delta_n]-k_n}\bgamma_{mt}h_{t,ml} |\bX_t) \right] .
$$
\end{ass}

 \begin{thm}[long-run g]\label{t5.5} Consider the known factor case 
Let $\mathcal P$ be the collection of all data generating processes $\mathbb P$ for which   Assumptions \ref{a:dgp}-\ref{a:gammabound:longrun} hold. 
Then for any fixed vector $\bv\in\mathbb R^{K}\backslash\{0\}$ such that $\|\bv\|>c>0$,  for each fixed $l\leq p$,
\begin{eqnarray*}
\sup_{\mathbb P\in\mathcal P}\left|\mathbb P\left(\bv'\int_0^T\bg_{lt}dt\in \widetilde{CI_{n,\tau}}\right)-( 1-\tau)\right|\to 0.
\end{eqnarray*}
 \end{thm}


\section{Models of Many Continuous-Time Moment Conditions}\label{s:gmm}

\subsection{The model}

We consider a  more general continuous-time model with linear moment conditions. 
Suppose we observe data  that are discretized  realizations from a continuous-time stochastic process $\{\bZ_t=(\bZ_{1t},...,\bZ_{pt}): t\in[0,T]\}$.  We assume $\bZ=\{\bZ_t\}_{t\geq 0}$ be a multivariate It\^{o} semimartingale on a filtered probability space $(\Omega, \calF, \{\calF_t\}_{t\geq0}, \bbP)$.  Igonoring jumps, we have 
$$
	 \bZ_{lt} =\int_0^t  \balpha_{ls}^Z ds + \int_0^t  \bsigma_{ls}^Z d \bW^Z_{ls}, \quad l=1,..., p,\quad \forall t\in[0,T]
$$
 where $\{ \bW_{lt}^Z\}_{t\leq 0}$ is a    Brownian motion, and $ \{ \balpha_{lt}^Z\}_{t\geq0}$ is the drift process. 
 For each $t\in[0, T],$ let the parameter $ \bbeta_{lt} $ be     identified by the following moment condition:
 \begin{equation}\label{e:4.1:moment:gener}
  \Psi_l(\bbeta_{lt},  \bc_{z, lt})  =0,\quad l=1,..., p,
 \end{equation}
where   $\bc_{z,lt}=d[\bZ_l,\bZ_l]_t/dt$ is  the instantaneous quadratic variation process of $\bZ_l=\{\bZ_{lt}\}_{t\geq 0}$, 
and  $ \Psi_l(\bbeta,\bc)$ is a known   function  linear with respect to $\bbeta$ and  continuously differentiable with respect to $\bc$. 
 In addition, we assume that $\bbeta_{lt}$ depends on  a set of characteristics $\bX_{lt}$, so has the following decomposition:
\begin{align} \label{e4.2}
   \bbeta_{lt} = \bg_{t}(\bX_{lt}) + \bgamma_{lt}, \quad l=1,\cdots,p,
\end{align}
where $\mathbb E(\bgamma_{lt}|\bX_{t})=0$.   Here $\bX_{t}$ a set of  (possibly) time-varying characteristics. The effect of characteristics on $\bbeta_{lt}$ is  represented by $ \bg_{t}(\bX_{lt}) $. The goal is to make uniform inference about $ \bg_{t}(\bX_{lt}) $ that allows the cross-sectional variations of $\bgamma_{lt}$ to be possibly arbitrarily close to the ``boundary".

Many applications in economics and finance give rise to the continuous-time \textit{many linear moment conditions}  as in 
(\ref{e:4.1:moment:gener}) and (\ref{e4.2}),  as we now illustrate with a few examples.

\begin{exm}[Multivariate regression models]
Consider a system of multivariate  continuous-time linear regression models with individual-specific regressors, e.g., \cite{barndorff2004econometric,mykland2006anova,kalnina2012nonparametric, li2017adaptive}:
\begin{gather}\label{e:hfregre}
\begin{split}  
    Y_{lt} =Y_{l0} &+ \int_0^t \alpha_{ls} ds + \int_0^t \bbeta_{ls}' d\bF_{ls} + U_{lt}:\quad l\leq p
\end{split}
\end{gather}
where we observe realizations of $(Y_{lt}, \bF_{lt})$ for $l=1,...,p.$
Assume that the quadratic covariation of $U_{lt}$ and $\bF_{lt}$ be zero, then we have 
$$
\Psi_l(\bbeta_{lt},  \bc_{z, lt}) =\bc_{FF,lt}\bbeta_{lt} - \bc_{YF,lt}
$$
with $(\bc_{FF,lt}, \bc_{YF,lt})$ respectively being the quadratic variation of the individual-specific regressor $\bF_{lt}$, and the quadratic covariation between  $Y_{lt}$ and $\bF_{lt}$. 
Here we consider a high-dimensional system  where $p\to\infty$ fast.  In particular, this model admits the linear factor model described in Section 3 as a special case, by setting $\bF_{lt}=\bF_t$ as a common factor, for all $l\leq p.$
 \end{exm}

\begin{exm}[Idioscynractic variance models]\label{ex4.2} Consider a continuous-time factor model 
$$
    Y_{lt} =Y_{l0}  + \int_0^t \alpha_{ls} ds + \int_0^t \btheta_{ls}' d\bF_{s} + U_{lt}:\quad l\leq p
$$
but we  use $\btheta_{lt}$ to denote the factor ``betas". Under the condition that $U_{lt}$ and $\bF_t$ are two  orthogonal processes, the factor betas can be identified as $\btheta_{lt}= \bc_{FF, t}^{-1}\bc_{YF, lt}$, and yield the following covariance decomposition: 
\begin{equation}\label{eidiosbeta}c_{YY, lt} =\btheta_{lt}'\bc_{FF, t} \btheta_{lt}+c_{uu,lt}
=\bc_{YF, lt}'   \bc_{FF, t}^{-1}\bc_{YF, lt}+c_{uu,lt}
\end{equation}
where $c_{YY, lt}$ and $c_{uu, lt}$ respectively denote the quadratic variations of $Y_{lt}$ and $U_{lt}$.  
Consider a model where  firms' idiosyncratic variances possess a high degree of comovement: the idiosyncratic variance $c_{uu,lt}$ is proportional to the market factor, with a  firm-specific scalar coefficient $\bbeta_{lt}\in\mathbb R$. That is,
\begin{equation}\label{e:4.4nointercep}
c_{uu,lt}=\bbeta_{lt}  c_{FF1,t} +\bar c_l,
\end{equation}
where $  c_{FF1,t}$ denotes the quadratic variation of the market factor,  and to  identify $\bbeta_{lt}$  we assume that $\bar c_l$ is a known  intercept in (\ref{e:4.4nointercep}).  The   model is in spirit  similar to the   moment restriction on the quadratic variations studied by \cite{li2016inference}\footnote{
\cite{li2016inference} 
allow   unknown intercepts in (\ref{e:4.4nointercep}). They are able to identify $\bbeta_l$ as it is assumed to be time-invariant over $t\in[0,T]$ and can be estimated using all sampling intervals on the entire time span. In contrast, we allow $\bbeta_{lt}$ to be time-varying, so to  ensure  $c_{uu,lt}>0$, we set $\bar c_l=\bar\beta_l\bar c_F$, where $\bar\beta_l$ and $\bar c_F$ are respectively   known upper bounds for $\max_{t\in[0,T]}|\bbeta_{lt}|$ and $\max_{t\in[0,T]}|c_{FF1,t}|$. Alternative,  one could allow an unknown intercept  $c_{uu,lt}=\bbeta_{lt}  c_{FF1,t}+c_{lt}$ and set another identification restriction, e.g., $a\bbeta_{lt}=c_{lt}$ for some known $a\in\mathbb R$. }.  Here $\bbeta_{lt}$   is referred to as the ``idiosyncratic variance
beta", and can be used to measure the relation between  the  idiosyncratic volatility for stock $l$ and the market factor  (e.g., \cite{ang2009high,herskovic2016common}). As evidenced by \cite{herskovic2016common},  households may face common fluctuations in their idiosyncratic variances.   Substituting (\ref{eidiosbeta}) to (\ref{e:4.4nointercep}), we obtain a moment condition, with 
 $$
 \Psi_l(\bbeta_{lt},  \bc_{z, lt})=\bbeta_{lt} c_{FF1,t}+\bar c_l+\bc_{YF, lt}'   \bc_{FF, t}^{-1}\bc_{YF, lt}-c_{YY, lt} .
 $$
 Then the beta decomposition (\ref{e4.2}) expresses the   idiosyncratic variance
beta into the sum of a characteristic-dependent component and  an orthogonal component.  For a  given individual $l$, we are interested in testing whether its idiosyncratic variance beta can   be explained by its characteristics, which corresponds to the null hypothesis that 
$ \bg_{t}(\bX_{lt})=0$ almost surely. 



 \end{exm}

\subsection{Estimation}\label{est:two-ste}
We apply the continuous-time generalized methods of moments (GMM)  based on the following moment conditions:
 \begin{eqnarray}\label{gmm}
  \Psi_l(\bbeta_{lt},  \bc_{z, lt}) &=&0,\quad \forall t\in[0, T], \cr
\mathbb E(\bbeta_{lt}|\bX_{lt})&=&\bg_{t}(\bX_{lt}),\quad l=1,..., p.
 \end{eqnarray}
  We shall first assume the underlying $\bZ_{lt}$ is continuous. 
  Let  $\dim(\bZ_{lt})= K_z$   and  $\dim(\Psi_l)=K_\psi$. We shall assume  $  K_z $ and $K_\psi$  are fixed constants, and $K_\psi\geq K$, so $\bbeta_{lt}$ is possibly over-identified.

  Ignoring the jumps, over the $i$-th sampling interval, we have the following discrete time observation:  
 \begin{align}\label{e:discretemodel}
 \begin{split}
 \Delta_i^n\bZ_l &= \bZ_{l,i\Delta_n} - \bZ_{l,(i-1)\Delta_n} = \int_{(i-1)\Delta_n}^{i\Delta_n} \Big( \balpha_{ls}^Z ds + \bsigma_{ls}^Z d \bW^Z_{ls}\Big) 
 \end{split}
 \end{align}
 Let $\widehat \bc_{z, lt}$ be the sample quadratic variation of $\bZ_{lt}$ on window $I_t^n$:
 $$
 \widehat \bc_{z, lt}=\frac{1}{k_n\Delta_n}\sum_{i\in I_t^n} \Delta_i^n\bZ_l \, \Delta_i^n\bZ_l'.
 $$
We employ a simple two-step estimation: 

\textit{Step 1.} 
Let $\widehat\bbeta_{lt}$ be the solution that satisfies:
\begin{equation}\label{zestimation}
\widehat\bbeta_{lt}:=\arg\min_{\bbeta}\Psi_l(\bbeta,  \widehat\bc_{z, lt})' \bOmega_{lt}  \Psi_l(\bbeta,  \widehat\bc_{z, lt})  ,\quad l=1,..., p.
\end{equation}
Let $\widehat\bbeta_{t}=(\widehat\bbeta_{1t},...,\widehat\bbeta_{pt})'$ be the $p\times K$ matrix. Here $\bOmega_{lt}$ is a known positive definite  weight matrix.  In the general case with jumps,  we solve
 $$
\widehat\bbeta_{lt}:=\arg\min_{\bbeta}\Psi_l(\bbeta,  \widehat\bc_{z, lt}^\psi)' \bOmega_{lt}  \Psi_l(\bbeta,  \widehat\bc_{z, lt}^\psi)  ,\quad l=1,..., p.
 $$ 
where  we replace each $\Delta_i^n \bZ_l$   with their truncated versions:
 $
 \widehat \bc_{z, lt}^{\psi}=\frac{1}{k_n}\sum_{i\in I_t^n} \Delta_i^n\bZ_l^{\psi_n} \, \Delta_i^n\bZ_l^{\psi_n\prime}.
 $
Here $\Delta_i^n \bZ_{l}^{\psi_n}:=\Delta_i^n \bZ_l \, 1_{\{ \| \Delta_i^n \bZ_l \| \leq \psi_n \}}$ denotes the usual truncated process for the process $\Delta_i^n \bZ_{l}$, with some random sequence $\psi_n$ that converges in probability to zero as $\Delta_n\rightarrow0$.

 \textit{Step 2.}     Define
 \begin{eqnarray*}
 \widehat\bG_t&=&(\widehat\bg_{1t},...,\widehat\bg_{pt})':=\bP_t\widehat\bbeta_{t},\cr
  \widehat\bGamma_t&=&(\widehat\bgamma_{1t},...,\widehat\bgamma_{pt})':=(\bI_p-\bP_t)\widehat\bbeta_{t}  .
\end{eqnarray*}

  
 Therefore, the estimator (\ref{e3.2}) in the linear factor model with known factors is a  special case of the two-step estimator presented here, while the estimator  of the unknown factor case can also be considered as a special case, which replaces $\Delta_i^n\bZ_l$ in the definition of $\widehat\bc_{z,lt}$ with  the ``estimated regressors".\footnote{ The general  asymptotic theories presented in this section focus on    the ``known  regressor case". In the case of ``estimated regressors", corresponding to the latent factor case in the linear factor model, the estimated regressors should be estimated separately and plugged in, as we did for the unknown factors in Section 3. The effect of estimating the unknown regressors might introduce additional biases that need be corrected. For simplicity, we do not   cover this case in the general GMM framework.}

\subsection{General theory for two-step estimations}\label{sec:gene:theo}

 Our  estimator belongs to the   general  class of two-step GMM estimators, as previously discussed in \cite{newey1994asymptotic, chen2003estimation,chen2015sieve,chernozhukov2016double}:  a nuisance parameter is estimated in the first step, and substituted in the second step estimation.
As  is shown by these authors, when the  second set of moment conditions is not ``Neyman orthogonal" with respect to $\bbeta_{lt}$ (roughly speaking, its ``directional derivative" with respect to $\bbeta_{lt}$ is nonzero), the first-step estimation error $\widehat\bbeta_{lt}-\bbeta_{lt}$ is not negligible, and plays a leading role in the   asymptotic distribution for $\widehat\bg_{lt}$. 

  In the current  large panel context with many moment conditions ($l=1,..., p$), indeed the second moment condition for $\bg_{lt}$, given by $\mathbb E(\bbeta_{lt}|\bX_t)=\bg_{lt}$, is not Neyman orthogonal with respect to $\bbeta_{lt}$.
However,    two new phenomena are present here. To see this, let $\nabla_\beta\Psi_m( \bc )$ be the  $K\times K$ gradient matrix of  $\Psi_m(\bbeta, \bc)$ with respect to $\bbeta$, which does not depend on $\bbeta$ due to the linearity.  Write
  \begin{eqnarray*}
  	\bA_{mt}(\bc):= (\nabla_\beta\Psi_m( \bc ) '\bOmega_{mt}\nabla_\beta\Psi_m( \bc )    )^{-1} \nabla_\beta\Psi_m( \bc)'\bOmega_{mt} .
  \end{eqnarray*}  
  The first-order condition of  (\ref{zestimation}) leads to, the $p\times K$ matrix $\widehat\bbeta_t$ satisfies:  
  $$
  \widehat\bbeta_t-\bbeta_t\approx-\begin{pmatrix}
\Psi_1(\bbeta_{1t}, \widehat\bc_{z,1t}) '    \bA_{1t}(\bc_{z,1t})'\cr
\vdots\cr
\Psi_p(\bbeta_{pt}, \widehat\bc_{z,pt}) '    \bA_{pt}(\bc_{z,pt})'
  \end{pmatrix}
  $$
 where $\bbeta_{lt}$ and $\bc_{z, lt}$ are evaluated at the true values. Then applying Step 2 of the estimation, combined with the delta-method, yields, for each $l\leq p$, 
 $$
 \widehat\bg_{lt}-\bg_{lt} = -\underbrace{\frac{1}{p}\sum_{m=1}^p \bA_{mt}(\bc_{z,mt})\nabla_c\Psi_m(\bc_{z, mt})\vecc(\widehat\bc_{z,mt}- \bc_{z,mt})h_{t,ml}}_{\text{first-step effect:} (\ba)}+\underbrace{\bGamma_t'\bP_{t,l}}_{\text{second-step effect:} (\bb)}+ o_P(\frac{1}{\sqrt{k_np}})
 $$
 where   $\nabla_c\Psi_m( \bc_{z, mt})$ is the gradient of $\Psi_m(\bbeta_{mt},\bc_{z, mt})$ with respect to $\vecc(\bc_{z,mt})$.


 The first new phenomenon is that  the  first-step effect ($\ba$) is  a cross-sectional average of the estimation errors for $\{\widehat\bbeta_{mt}-\bbeta_{mt}:  m=1,..., p\}$, whose rate of convergence is $O_P((k_np)^{-1/2})$  and is   in fact negligible  when $\Var(\bGamma_t|\bX_{lt})$ is bounded away from zero, even though the second moment condition is not Neyman orthogonal.  This is essentially due to the fact that in the presence of many moment conditions,  the first-step effect     can be ``cross-sectionally averaged out"  and thus can be dominated by the second-step effects when the latter has a slower rate of convergence. 
 
The second new phenomenon,  which is also the unique feature in our model, is that  whether the second-step moment condition $\bE(\bbeta_{lt}|\bX_{t})= \bg_{lt}$ is  ``noise-free" is unknown, in the sense that it is possible that the magnitude of the eigenvalues of  $\Var(\bGamma_t|\bX_{lt})   $  can be arbitrary in their parameter space $[0, C]$, and may vary across time. Especially,  when $\Var(\bGamma_t|\bX_{t}) $ is near the boundary of the parameter space (zero), which occurs when $\bX_{t}$ has nearly full explanatory power on $\bbeta_{lt}$, the first-step effect becomes the only leading term in the asymptotic expansion. 

 The aforementioned two unique features of our model call for a different asymptotic analysis for the two-step estimator considered here, and similar to the discussions in the linear factor model, lead to an unknown rate of convergence and a discontinuity in the limiting distribution.

To formally present our theory, we assume that   the following conditions hold uniformly over a class of DPG's: $\mathbb P\in\mathcal P$.

\begin{ass}\label{adgp:general}

(i) For each $l\leq p$, $\bZ_{lt}$ is a multivariate It\^{o} semimartingale on a filtered probability space $(\Omega, \calF, \{\calF_t\}_{t\geq0}, \bbP)$, whose  continuous component is given by 
$$
	\bZ_{lt}^c= \int_0^t  \balpha_{ls}^Z ds + \int_0^t  \bsigma_{ls}^Z d \bW^Z_{ls}, \quad \forall t\in[0,T]
$$
 where $\{ \bW_{lt}^Z\}_{t\leq 0}$ is a    Brownian motion, and $ \{ \balpha_{lt}^Z\}_{t\geq0}$ is the drift process.  The jump components of $\bZ_{lt}$ are of finite variation uniformly over $l\leq p$.

(ii) Write  $\btheta_{lt}:= ( \bX_{lt}, \bgamma_{lt}, \balpha_{lt}^Z, \bsigma_{lt}^Z)$. For each $l\leq p$, $\btheta_{lt}$ is a multivariate continuous and locally bounded (uniform in $l\leq p$) It\^{o} semimartingale with the form: 
\begin{align*}
	\btheta_{lt} =\, & \btheta_{l0} + \int_0^t \tilde{\balpha}_{ls} ds + \int_0^t \tilde{\bsigma}_{ls} d\bW_s^l .
\end{align*}
Here $ \{\widetilde\balpha_{ls}\}_{s\geq0}$ and $ \{\widetilde{\bsigma}_{ls}\}_{s\geq0}$  are optional processes and locally bounded uniformly in $l\leq p$.  We allow $\bW_t^l$ to be correlated with $\bW_{lt}^Z$.

(iii) $\E(\bgamma_{lt}|\bX_{t})=0$ for all $t\in[0, T), l\leq p$.

\end{ass}

Next,  for each $m\leq p$,  $\Psi_m(\bbeta, \bc)$ is linear in $\bbeta$, so $\nabla_\beta\Psi(\bc)$ does not depend on $\bbeta$. When $\bbeta_{mt}$ is evaluated at  the true value, 
we simply write $ \Psi_m(   \bc ):=  \Psi_m(\bbeta_{mt},   \bc )$ and $\nabla_c\Psi_m(   \bc ):=\nabla_c\Psi_m(\bbeta_{mt},  \bc )$
to explicitly make them as functions of $\bc$. Here $\nabla_\beta,\nabla_c$ respectively denote the gradient operators with respect to $\bbeta$ and $\vecc(\bc)$.  
   \begin{ass} 
 \label{ahigh1} 
 (i) $\Psi_m(\cdot,\cdot)$ is twice continuously differentiable for all $m\leq p$.

(ii) $\max_{m\leq p}\|(\nabla_\beta\Psi_m( \bc_{z,mt}  ) '\bOmega_{mt}\nabla_\beta\Psi_m( \bc_{z,mt} ) )^{-1}\| {=O_P(1)}$, $\max_{m\leq p}\|\bOmega_{mt}\|=O_P(1)$.

 \end{ass}

 \begin{ass}The sample quadratic variation satisfies:  at each fixed $t\in[0,T]$, 
 \label{ahigh0} 
  \begin{eqnarray*}
(i)&& \frac{1}{p}\sum_{m=1}^ph_{t,ml} \bA_{mt}( \widehat \bc_{z,mt}  ) [ \Psi_m( \widehat \bc_{z,mt}) - \Psi_m(\bc_{z,mt})  - \nabla_c\Psi_m(  \bc_{z,mt}) \vecc(\widehat \bc_{z,mt}- \bc_{z,mt} )] =o_P\left(\frac{1}{\sqrt{pk_n}}\right) \cr
(ii) &&\frac{1}{p}\sum_{m=1}^ph_{t,ml}  [   \bA_{mt}(\widehat \bc_{z,mt})    -  \bA_{mt}(  \bc_{z,mt})  ]   
\nabla_c\Psi_m( \bc_{z,mt}) \vecc(\widehat \bc_{z,mt}- \bc_{z,mt} ) =o_P\left(\frac{1}{\sqrt{pk_n}}\right)    .
 \end{eqnarray*}
 \end{ass}
 Assumption \ref{ahigh0} is a high-level assumption on the effect of estimating a large number of  quadratic variations. It requires two fundamental conditions. The first line requires that  the moment function $\Psi_m( \bc)$ should be well approximated by linear functions locally; the second line additionally requires the smoothness of $ \bA_{mt}(\bc)$. 
A loose  upper bound using   $\frac{1}{p}\sum_{m=1}^p \|\widehat\bc_{z,mt}-\bc_{z,mt}\|^2=O_P(k_n^{-1})$ based on the Cauchy-Schwarz inequality can be achieved,    so both conditions can be simply verified so long as $p=o(k_n)$.  On the other hand, sharper upper bounds can also be achieved to allow for a much larger $p$, by noting that both conditions 
  are regarding high-dimensional cross-sectional averages  of the effects of $\widehat \bc_{z,mt}- \bc_{z,mt}$, and intuitively, these   should be ``averaged out" as $p\to\infty, $    
 and  can be verified case-by-case for  specific models. 
For instance, \cite{powerenhancement} and \cite{bai2017inferences} provide   technical arguments to verify (ii) when $\Psi_m(\bc)$ is   linear in   $\bc^{-1}$.   In fact, when proving Theorem \ref{t:g} in the linear factor model, we verify these conditions directly and apply our general result  theorem \ref{t:general}.


We now describe the asymptotic variance of $\widehat \bg_{lt}$.  For $m=1,..., p$, let 
  $$
  \bxi_{mt,s}:=\bA_{mt}(\bc_{z,mt})\nabla_c\Psi_m(  \bc_{z, mt})\vecc\left(\frac{1}{s}(\bZ_{m,t+s}-\bZ_{mt})  (\bZ_{m,t+s}-\bZ_{mt})'-\bc_{z,mt }   \right).
  $$
  
   The asymptotic distribution of $\widehat\bg_{lt}$ is jointly determined by $\bxi_{mt,s}$ and $\bgamma_{mt}$.   Let 
\begin{eqnarray}\label{ecov:genera}
\bV_{u, t} &=&  \lim_{s\to0}\Var\left( \frac{1}{ \sqrt{p}}\sum_{m=1}^p h_{t,ml}  \bxi_{mt,s}\bigg{|}\mathcal F_t\right)   \cr
\bV_{\gamma, t} &=& \Var\left(\frac{1}{\sqrt{p}}\sum_{m=1}^ph_{t,ml}\bgamma_{mt}\bigg{|}\bX_t\right).
\end{eqnarray}

\begin{ass} \label{a:errorcov4.6}  

  There are  absolute constants $c, C>0$,  almost surely, 
  
  (i) $\max_{m\leq p}\sup_{t\in[0, T]}\Var(\bgamma_{mt}|\bX_t)<C$, and 
$$
c< \inf_{t\in[0,T]}\lambda_{\min}(\bV_{u,t})\leq  \sup_{t\in[0,T]}\lambda_{\max}(\bV_{u,t})<  C,\qquad 
 \lambda_{\max}(\bV_{\gamma,t}) \leq  C\lambda_{\min}(\bV_{\gamma,t}).$$

  (ii)   $\sup_{t\in[0, T]}\max_{s\in[0,\Delta_n]} \mathbb E(\|\frac{1}{\sqrt{p}}\sum_{m=1}^p h_{t,ml} \bxi_{mt,s}\|^4 |\mathcal F_{t})=O_P(1) $.

\end{ass}

 Generally, we have the following theorem. 
   
\begin{thm} \label{t:general} Suppose $J, p, k_n\to\infty$ and $\Delta_n\to0$ satisfy: $J^2=O(p)$,   $k_npJ^{-2\eta} +pk_n^2\Delta_n =o(1)$.  
 Under Assumptions   \ref{a3.2:smooth:g}, \ref{a:sieve}, \ref{adgp:general}-\ref{a:errorcov4.6},  $$
\left(\frac{1}{k_np}\bV_{u,t}+\frac{1}{p} \bV_{\gamma, t}\right)^{-1/2}\big(\widehat \bg_{lt}-\bg_{lt} \big) \ConStable N(0,\bI_K).
$$
\end{thm}

Therefore, when the strength of $\bV_{\gamma, t}$ is unknown, the general two-step GMM estimation in the current context yields an unknown rate of convergence $O_P(a_{np})$, where $a_{np}$ may vary on the range $[(pk_n)^{-1/2}, p^{-1/2}]$. This leads to 
 a discontinuity on its limiting distribution. 

 \begin{remark}
The weight matrices $\{\bOmega_{mt}: m\leq p\}$ are   involved in the asymptotic distribution through $\bV_{u,t}$. Similar to the usual GMM setting, the optimal weight matrix can be determined by   optimizing $\bV_{u,t}$, and is given by
$$
\bOmega_{mt} =\left(\nabla_c\Psi_m(  \bc_{z, mt}) \bV_{mt} \nabla_c\Psi_m(  \bc_{z, mt})' \right)^{-1},
$$
where 
$ \bV_{mt}=\Var\left[\vecc\left(\frac{1}{\Delta_n}\Delta_i^n\bZ_{m}\Delta_i^n\bZ_{m}' \right)\bigg{|}\mathcal F_t\right].$ In the case of exact identification, as in the linear factor models, the weight matrix $\bOmega_{mt}$ no longer plays a role in the asymptotic distribution. 
 \end{remark}

\subsection{Cross-sectional Bootstrap}\label{sec:boot:gene}

We extend the cross-sectional bootstrap described in Section 3  to the more general context, in order to  achieve the uniform inference. Importantly, note that the main issue that causes the unknown limiting distribution is  from 
$  {\bGamma_t'\bP_{t,l}} $ in the expansion of $\widehat\bg_{lt}-\bg_{lt}$. This term comes from the second-step estimation. Hence    the first-step regressions for $\bbeta_{t}$, which only depends on time-series estimations, is not needed to be repeated in the bootstrap steps. 

As such,   we directly resample    $\{ (\widehat\bbeta_{mt}^*, \bX_{mt}^*):  m=1,... , p \}$ with replacement, where 
\begin{eqnarray*}
\{ \widehat\bbeta_{mt}^*, \bX_{mt}^*:  m=1,... , p \}
=\big\{ (\widehat\bbeta_{m_1, t}, \bX_{m_1, t}), ..., (\widehat\bbeta_{m_p, t}, \bX_{m_p, t})  \big\}
\end{eqnarray*}
 and   $\widehat\bbeta_{mt}$ is the estimated $\bbeta_{mt}$ in step 1.
 Here $\{m_1,..., m_p\}$ is a simple random sample with replacement from $\{1,..., p\}$.   
 We then let   $\bPhi_t^*=(\bphi_{m_1,t},...,\bphi_{m_p,t})'$, $\bP_t^*=\bPhi_t^*(\bPhi_t^{*'}\bPhi_t^*)^{-1}\bPhi_t^{*'}$,  and $\widehat\bbeta_t^*=(\widehat\bbeta_{m_1, t}^*,..., \widehat\bbeta_{m_p, t}^*)'. $
 When we are interested in $\bg_{lt}$ for given $l$, we always 
fix the index of the $l$-th resampled cross-sectional unit being $l$, that is, $m_l=l$. Hence $\widehat\bbeta_{lt}^*=\widehat\bbeta_{lt}$. 
 Define
 $$
 \widehat\bG_t^*= \bP_t^*\widehat\bbeta_{t}^*
 $$
 and let $ \widehat \bg_{lt}^*$ be the $l$-th column  of $ (\widehat\bG_t^*)'$.

When $\bg_{lt}$ is multidimensional,  we present the confidence interval for a linear transformation   $\bv'\bg_{lt}$.
 
\begin{algo} Compute the confidence interval for   $\bv'\bg_{lt}$ as follows.
	
	Step 1.  Take a simple random sample $\{m_1,...,m_p\}$ with replacement from $\{1,...,p\}$.  Fix $m_l=l$. 
	
	Step 2. Compute $ \widehat \bg_{lt}^{*}$  as  described above. 
		
	Step 3. Repeat Step 1-2 for $B$ times  and obtain   $\{\widehat \bg_{lt}^{*b}\}_{b\leq B}$. For a predetermined confidence level $1-\tau$, let  $q_{\tau}$ be the   $1-\tau$ bootstrap quantile of $\{  |  \bv'\widehat \bg^{*b}_{lt} -   \bv'\widehat \bg_{lt} |  \}_{b\leq B}$. 
	  
	  Step 4. Compute the confidence interval as:
	 \begin{eqnarray*}
	 	CI_{nt, \tau}&=&[\bv'\widehat \bg_{lt}  - q_\tau, \bv'\widehat \bg_{lt} +q_\tau]. 	 \end{eqnarray*}
\end{algo}

For the first-order validity of the bootstrap, we  impose a  high-level assumption as Assumption \ref{ahigh0} in the bootstrap sampling space. In this assumption,  $ \{\bA_{mt}^*(\bc), \Psi_m ^*(\bc), \widehat \bc_{z,mt} ^*,  \bc_{z,mt} ^* : m\leq p\}$
denote the  bootstrap samples of  $ \{\bA_{mt}(\bc), \Psi_m (\bc), \widehat \bc_{z,mt} , \bc_{z,mt}  : m\leq p\}$;  and $h_{t,ml}^*=\bphi_{lt}^{*'}(\frac{1}{p}\bPhi_t^{*'}\bPhi_t^*)^{-1}\bphi_{mt}^*$.
 
 \begin{ass} \label{ahigh0:boot}  At each fixed $t\in[0,T]$, 
  \begin{eqnarray*}
(i)&& \frac{1}{p}\sum_{m=1}^ph_{t,ml} ^*  \bA_{mt} ^*( \widehat \bc_{z,mt}  ^* ) [ \Psi_m ^*( \widehat \bc_{z,mt} ^*) - \Psi_m ^*(\bc_{z,mt} ^*)  - \nabla_c\Psi_m ^*(  \bc_{z,mt} ^*) \vecc(\widehat \bc_{z,mt} ^*- \bc_{z,mt} ^* )] =o_{P^*}\left(\frac{1}{\sqrt{pk_n}}\right) \cr
(ii) &&\frac{1}{p}\sum_{m=1}^ph_{t,ml} ^*  [   \bA_{mt} ^*(\widehat \bc_{z,mt} ^*)    -  \bA_{mt} ^*(  \bc_{z,mt} ^*)  ]   
\nabla_c\Psi_m ^*( \bc_{z,mt} ^*) \vecc(\widehat \bc_{z,mt} ^*- \bc_{z,mt}  ^*) =o_{P^*}\left(\frac{1}{\sqrt{pk_n}}\right)    .
 \end{eqnarray*}
 \end{ass}

Next, recall that   $
\bxi_{mt,s}:=\bA(\bc_{z,mt})\nabla_c\Psi_m(  \bc_{z, mt})\vecc\left(\frac{1}{s}(\bZ_{m,t+s}-\bZ_{mt})  (\bZ_{m,t+s}-\bZ_{mt})'-\bc_{z,mt }   \right).
$

\begin{ass} \label{av}  
For each $t\in[0,T]$ and $s\in[0, \Delta_n]$, there exist    $\{\bxi_{mt,s}^0: m\leq p \}$ that satisfies:

(i)  Conditionally on $\bX_t$, $\{(\bgamma_{mt}, \bxi_{mt,s}^0): m\leq p\}$ are cross-sectionally uncorrelated.

  (ii)   $
\sup_{t\in[0,T]}\max_{s\in[0,\Delta_n]}  \frac{1}{p} \sum_{m\leq p}\EE(\|\bxi_{mt,s}-\bxi_{mt,s}^0\|^4|\mathcal F_{t})=O_P((pk_n)^{-2}).
  $

\end{ass}

We would like to emphasize that  Assumption \ref{av} does \textit{not} assume that  $\{\bZ_{mt}: m\leq p\}$ should be  cross-sectionally uncorrelated, because in  many applications it doe not hold for $\bZ_{mt}$, but in fact holds for $\bxi_{mt}^0$. In the linear regression model (\ref{e:hfregre}) for instance, 
 $\bZ_{mt}=(Y_{mt}, \bF_{mt})$, $m\leq p$, then $\{Y_{mt}: m\leq p\}$ would not be  uncorrelated if $\bF_{mt}$ contains common regressors (e.g., factors). On the other hand, it can be directly verfied that in this case,
  \begin{eqnarray*}
 	\bxi_{mt,s}^0&=&-\bc_{FF,mt}^{-1}\frac{1}{s} (\bF_{m,t+s} -\bF_{mt})  (U_{m,t+s} -U_{mt}) 
 	\cr
 	\bxi_{mt,s}&=&	\bxi_{mt,s}^0+\bdelta_{mt,s},
 \end{eqnarray*}
 where 
  $\bdelta_{mt,s} $ satisfies Assumption \ref{av} (ii), so is negligible;
 $\bc_{FF,mt}$ denotes the quadratic variation of $\bF_{mt}$. In the idiosyncratic volatility model (Example \ref{ex4.2}), it is also straightforward to verify that 
 $$
\bxi_{mt,s}^0=-\bc_{FF,t}^{-1} \left(\frac{1}{s}(U_{m,t+s}-U_{mt})^2 -c_{uu,mt}\right)h_{t,ml}.
$$
So $\bxi_{mt,s}^0$ are cross-sectionally uncorrelated given that $\{U_{m,t+s}-U_{mt}: m\leq p\}$ are cross-sectionally uncorrelated.

 \begin{thm}[Uniformly valid confidence intervals]\label{t:conf:boot:gene} Suppose $J^2=o(p)$, $J^{-\eta}pk_n+pk_n^2\Delta_n=o(1)$. 
Let $\mathcal P$ be the collection of all data generating processes $\mathbb P$ for which   Assumptions  \ref{a3.2:smooth:g}, \ref{a:sieve}, \ref{adgp:general}- \ref{av}   hold.    Then for any fixed vector $\bv\in\mathbb R^{K}\backslash\{0\}$  such that $\|\bv\|>c>0$,  for each fixed $l\leq p, t\in[0, T]$ 
\begin{eqnarray*}
 \sup_{\mathbb P\in\mathcal P}\left| \mathbb P(\bv'\bg_{lt}\in CI_{nt,\tau})-( 1-\tau)\right|\to0.
\end{eqnarray*}
 \end{thm}

\section{Empirical Studies}\label{sec: realdata}

\subsection{The data}

We use the  price data of stocks from the S\&P 500 index constituents for the period from July 2006 through June 2013.   We collect intraday transactions data of each stock from the TAQ database and construct returns every five minutes. We drop the overnight returns for excluding stock splits and dividend issuances, and abnormal prices that bounced back within a few seconds.  Stocks with missing price data are also dropped.  Therefore there are in total 380 stocks in our dataset.  In addition,  we construct the Fama-French four factors with five-minute frequency by  first generating the  five-minute returns of each common stocks on the NYSE, the AMEX, and the NASDAQ in the CRSP database and then following the method described in  \cite{FF}.    These  factors are: the market factor (Mkt), the small-minus-big market capitalization (SMB) factor,   high-minus-low book to market ratio (HML) factor, and the    profitability factor (RMW),  the difference between the returns of firms with high and low operating profitability.

We also collect fundamentals of those stocks from the Compustat database over the same period to construct firm characteristics. We consider four characteristics for each stock: size, value, momentum, and volatility as in \cite{CMO}. The annual size and value characteristic of each stock is the logarithm of the market value and the ratio of the market value to the book value in the previous June respectively. The monthly momentum and volatility characteristic of each stock is the cumulative returns of the last twelve months including the previous month, and the standard deviation of the last twelve months, including previous month respectively. 

\subsection{Cross-sectional variations}

\begin{figure}[htb!]
\centering
  \includegraphics[width=1\textwidth]{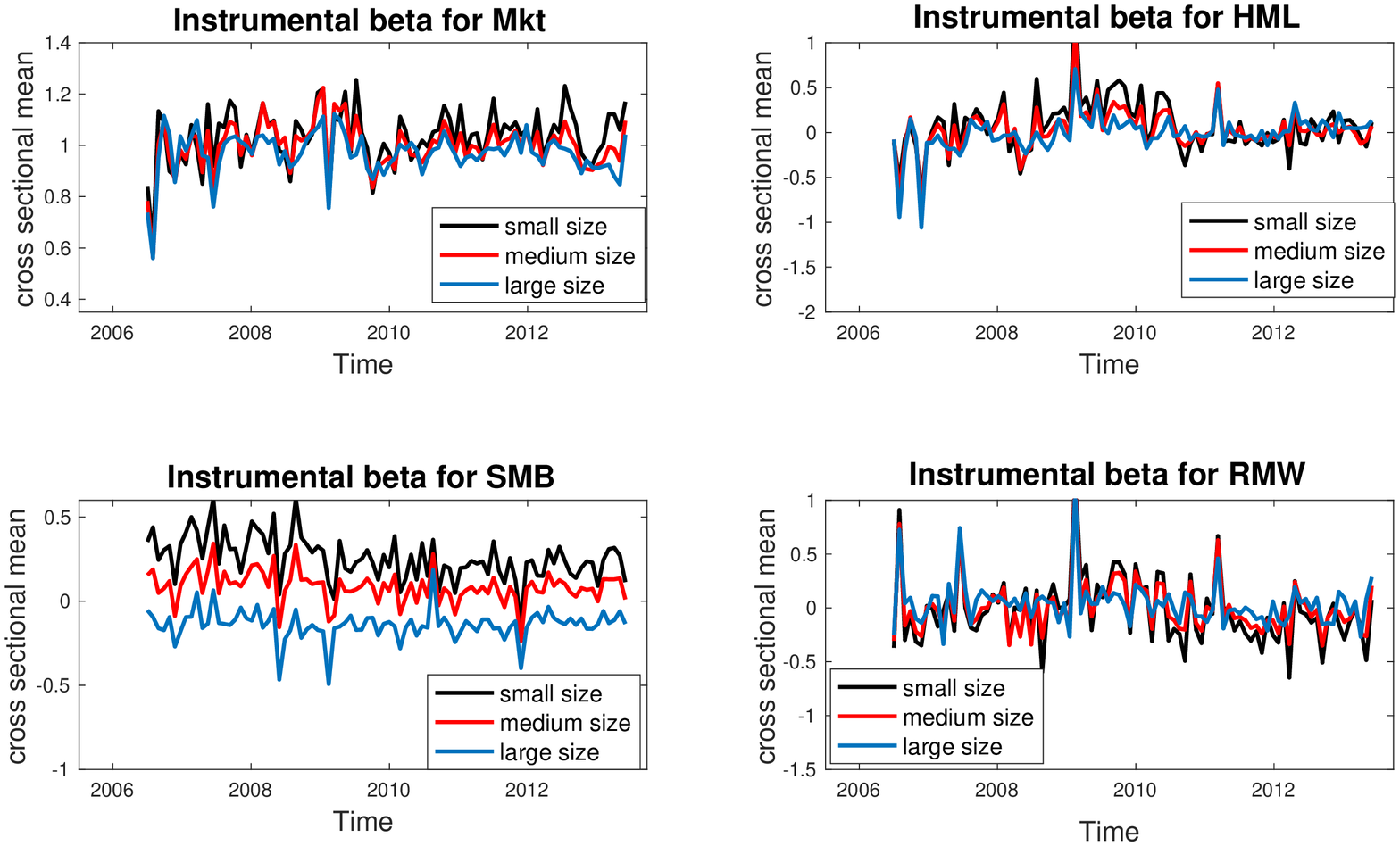}
   \includegraphics[width=1\textwidth]{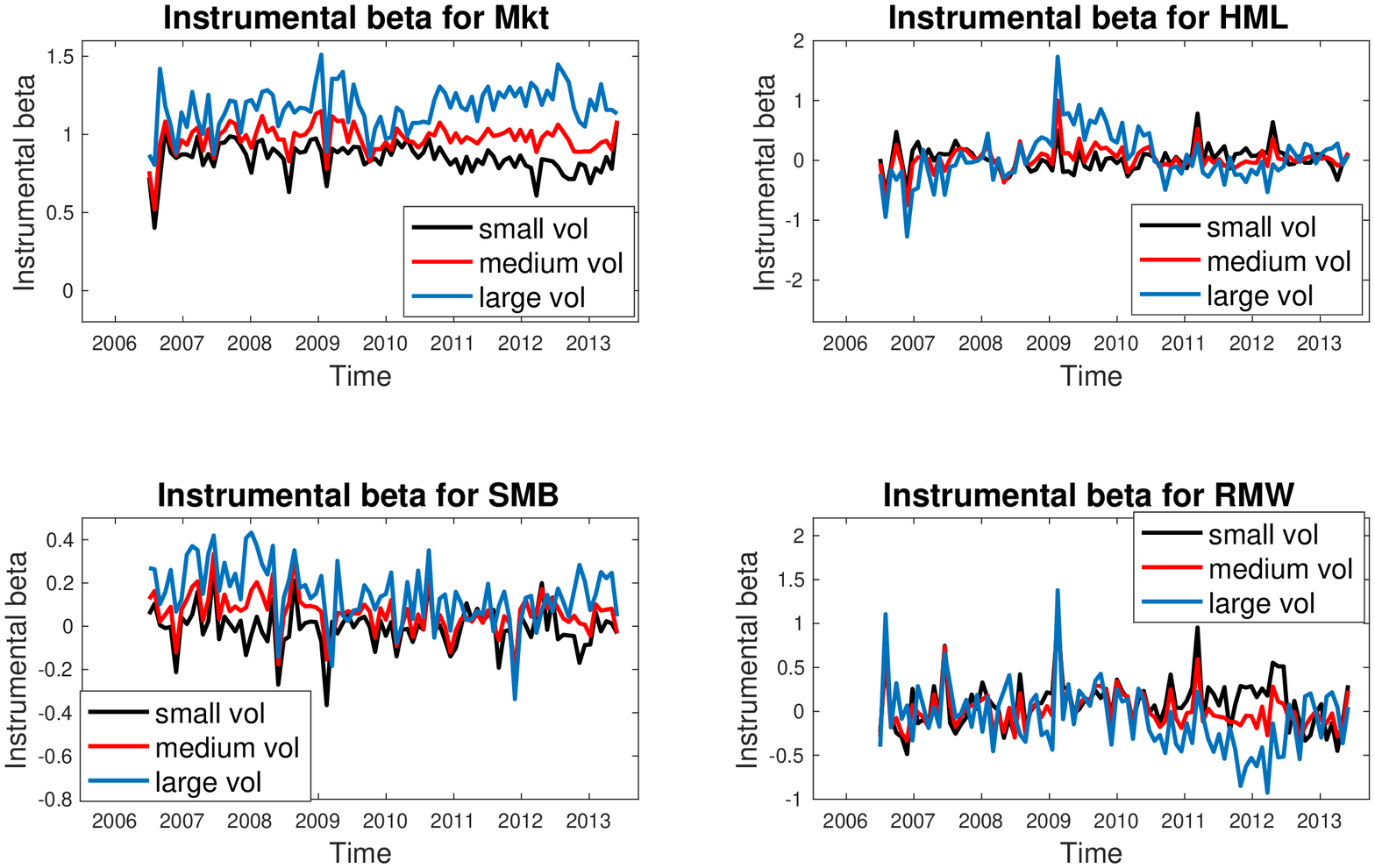}
\caption{\footnotesize Cross-sectional means of characteristic beta, grouped by size (upper four) and by volatility (lower four). The characteristic betas are estimated on a daily basis, and this figure plots eight days' estimations for each month.
}
\label{fig:1}
\end{figure}

Our estimation is based on the   5-minute frequency, and    intervals are taken as daily windows. Hence there are $k_n=78$ observations for each stock each day. We use the linear sieve basis, which are simply the standardized values of the four characteristics, and estimate the spot characteristic and idiosyncratic beta for each company on each trading day.\footnote{We also tried B-splines with degree 3 \citep{eilers1996flexible}, particularly for estimating and plotting the $\bg_t(\cdot)$ functions. Results obtained a very similar.} We divide the assets into three categories: large, medium, and low, based on  either the firms'  size or the volatility   characteristics. 
 Figure  \ref{fig:1} plots the cross-sectional average  of the characteristic betas corresponding to each of the four factors, classified by either the size or the volatility. Both size and volatility have noticeable effects on at least one of the factor betas.  The cross-sectionally averaged characteristic betas for the SMB factor are noticeably different across  three size groups, and in the long run, companies with larger size (market value) tend to be less sensitive to the SMB factor than companies with smaller size.  As shown in the fifth panel of Figure  \ref{fig:1},  companies with smaller volatilities   tend to be less sensitive to the market factor than companies with larger volatilities. While both phenomena have been documented in the literature, 
  the characteristic betas, however,  capture  long-run movements in beta driven by structural changes in the economic environment and in firm- or industry-specific conditions, so demonstrate long-run patterns in betas from these figures.

\begin{figure}[htb!]
\centering
  \includegraphics[width=1\textwidth]{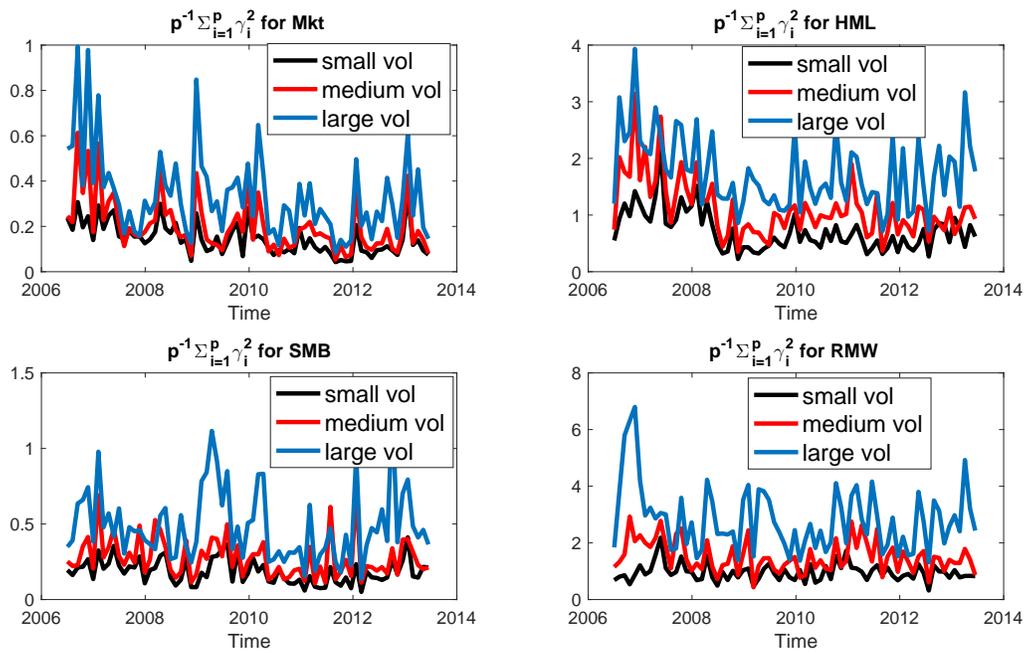}
\caption{\footnotesize Cross-sectional variations of $\bGamma$ across times. Firms are grouped into small, medium and large sizes. 
}
\label{fig:gammad}
\end{figure}

In addition, we also estimate the cross-sectional variations in the idiosyncratic betas, measured by $\frac{1}{|G|}\sum_{j\in G}\widehat \gamma_{lt,k}^2$ for $k=$ Mkt, HML, SMB and RMW factors. In the lower panel firms are grouped by volatility: $G\in\{\text{small vol}, \text{medium vol},  \text{large vol} \}$. So the computed measure shows the cross-sectional variations among firms of  small, medium and large volatilities. Figure \ref{fig:gammad} plots $\frac{1}{|G|}\sum_{j\in G}\widehat \gamma_{lt,k}^2$ over time. There are substantial differences on the cross-sectional variations among firms with difference sizes. In particular, the strength  of $\bGamma$  is the strongest and   also the most volatile across time for firms of large volatilities, is the weakest but least volatile across time for   firms of small volatilities. This measures different prediction power of the characteristics on betas among  firms of different level of volatilities. Figure \ref{fig:gammad}  also demonstrates that the strength of $\bGamma$ can be represented by various asymptotic sequences across time, so the uniform inference is very essential.

\begin{figure}[htb!]
\centering
  \includegraphics[width=16cm]{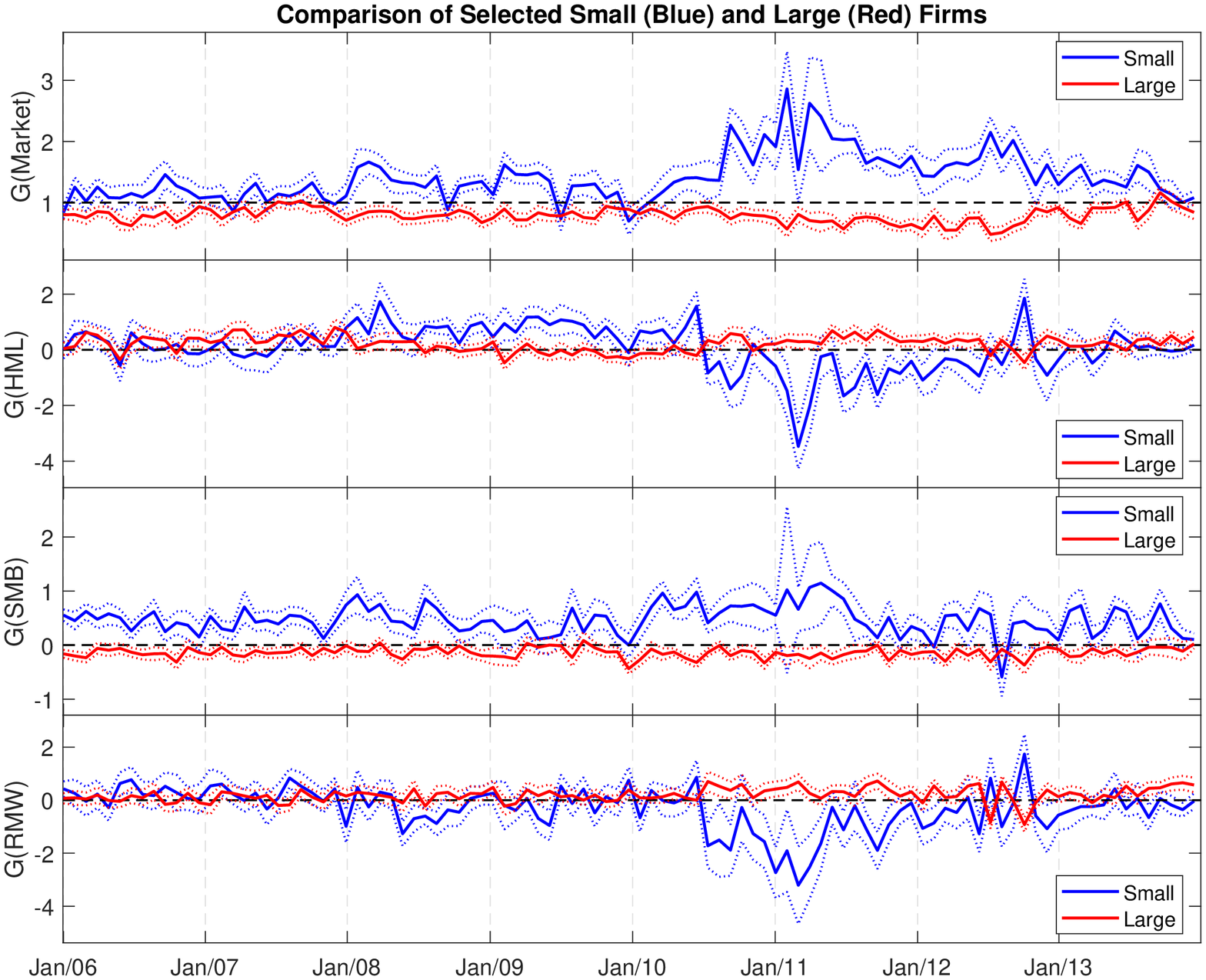}
\caption{\footnotesize Two individual stocks' confidence intervals: the two firms with the largest and smallest sizes in the dataset.
}
\label{fig:3}
\end{figure}


\normalsize


\subsection{Confidence Intervals}
We construct 95\% construct confidence intervals for each of the firms'   characteristic betas on a daily base, and   report and compare them among three groups (by either size or volatility).    On each trading day we construct the  confidence intervals and calculate the proportion of positive/negative significances among firms in each group. Then we average these (cross-sectional) proportions over all days within a fixed year, leading to the ``averaged proportion of significance" for each group. 

When the groups are formed by size, Table \ref{t2} reports the  results  of 2006, and we find that results of other years (2007 through 2012) demonstrate similar patterns: (1)  All stocks have significantly positive characteristic betas loading on the market factor. In fact, most of the characteristic betas for the market factor are larger than one. (2) There is a substantial difference in  the characteristic betas on  the SMB factor between firms of  small/medium size and firms of large size. Only 4.7\% of firms of large size have positive  significance, but this proportion is as high as 87\% for firms of small size. On the other hand, more than fifty percent of firms of large size have negative significance, but there are less than one percent of firms of small size. This shows that  the in-firm conditions and characteristics   produce a long-run mechanism making small firms   positively exposed and large firms negatively exposed   to the SMB systematic risk.
It becomes more interesting when we compare the results with the  proportions of $\bGamma$ and $\bbeta$. We find that for SMB, the proportion of positive $\bbeta$ is 37\% for large firms, and 71\% for small firms, while the proportion of negative $\bbeta$ is 62\%  for large firms, and 28\% for small firms. In contrast, these proportions    respectively become  51\% and   48\% for positive $\bGamma$, and 52\% for negative $\bGamma$, so the difference among firms of large and small sizes in $\bGamma$ is much less noticeable. This suggests that the characteristic beta is the main driving horse to determine the sign of $\bbeta$,  while the idiosyncratic beta is more related to beta's cross-sectional variations.  
 (3) As the size becomes larger, there is also a decreasing pattern on the negative significance of the  HML beta (more noticeable   on the SMB betas).
 \begin{table}[htp]
  \small
\caption{Cross-sectional Proportion of significant $\bG $ of groups by size, 2006 }
\begin{center}
\begin{tabular}{c|ccccccccc}
\hline
\hline
 &   \multicolumn{4}{c}{positive significance} &&   \multicolumn{4}{c}{negative significance} \\
 size & Mkt & HML & SMB & RMW &  & Mkt & HML & SMB & RMW \\
 \hline
small & 1 & 0.261 & 0.870 & 0.134 &  & 0 & 0.177 & 0.004 & 0.184 \\
medium & 1 & 0.234 & 0.450 & 0.162 &  & 0 & 0.215 & 0.039 & 0.150 \\
large & 1 & 0.133 & 0.047 & 0.154 &  & 0 & 0.229 & 0.544 & 0.062 \\
\hline
\end{tabular}
\end{center}
\label{t2}
\end{table}%
 
When we group  firms  by the volatility, however,  the pattern   demonstrates noticeable variations over years.  The results are given in Table \ref{t3}. Results of 2010 are similar to 2011, and results in 2007 are similar to 2006 so are not presented.   Firms with larger volatility tend to be more positively exposed to the HML factors than firms with smaller  volatility, who are more negatively exposed to HML.  This pattern appears in 2006, 2007,  2010 and 2011, but is reversed during the crisis period in 2008-2009, and European debt crisis 2012.

\begin{table}[htp]
\small
\caption{Cross-sectional Proportion of significant $\bG $ of groups by  volatility }
\begin{center}
\begin{tabular}{c|ccccccccc}
\hline
\hline
 &   \multicolumn{4}{c}{positive significance} &&   \multicolumn{4}{c}{negative significance} \\
volatility & Mkt & HML & SMB & RMW &  & Mkt & HML & SMB & RMW \\
\hline
   &   \multicolumn{9}{c}{2006}\\
small & 1 & 0.409 & 0.313 & 0.158 &  & 0 & 0.055 & 0.307 & 0.168 \\
medium & 1 & 0.187 & 0.473 & 0.151 &  & 0 & 0.159 & 0.214 & 0.153 \\
large & 1 & 0.034 & 0.583 & 0.142 &  & 0 & 0.403 & 0.068 & 0.075 \\
   &   \multicolumn{9}{c}{2008}\\
small & 1 & 0.180 & 0.320 & 0.313 &  & 0 & 0.400 & 0.318 & 0.049 \\
medium & 1 & 0.378 & 0.511 & 0.280 &  & 0 & 0.170 & 0.178 & 0.092 \\
large & 1 & 0.644 & 0.565 & 0.195 &  & 0 & 0.079 & 0.090 & 0.138 \\
   &   \multicolumn{9}{c}{2009}\\
small & 1 & 0.234 & 0.202 & 0.131 &  & 0 & 0.333 & 0.387 & 0.242 \\
medium & 1 & 0.286 & 0.341 & 0.152 &  & 0 & 0.296 & 0.230 & 0.223 \\
large & 1 & 0.346 & 0.506 & 0.184 &  & 0 & 0.198 & 0.086 & 0.171 \\
   &   \multicolumn{9}{c}{2011}\\
small & 1 & 0.397 & 0.285 & 0.521 &  & 0 & 0.164 & 0.295 & 0.050 \\
medium & 1 & 0.284 & 0.340 & 0.144 &  & 0 & 0.315 & 0.273 & 0.232 \\
large & 1 & 0.175 & 0.278 & 0.017 &  & 0 & 0.445 & 0.224 & 0.621 \\
   &   \multicolumn{9}{c}{2012}\\
small & 1 & 0.234 & 0.202 & 0.131 &  & 0 & 0.333 & 0.387 & 0.242 \\
medium & 1 & 0.286 & 0.341 & 0.152 &  & 0 & 0.296 & 0.230 & 0.223 \\
large & 1 & 0.346 & 0.506 & 0.184 &  & 0 & 0.198 & 0.086 & 0.171 \\

\hline
\end{tabular}
\end{center}
\label{t3}
\end{table}%

 We now focus on two individual stocks' confidence intervals. We take the two firms that  have  the highest frequency to be respectively classified in the ``large  group'' and the ``small group'' by size, and call them ``large'' and ``small''.   
 Figure \ref{fig:3} plots the estimated characteristic betas and the associated confidence intervals of the two firms over time. As for the beta associated with the market factor, while both are positively significant, the characteristic betas of the firm with smaller size are constantly larger than one, making it more sensitive to the changes of market risks than the firm with the larger size. In addition, the pattern shown by the characteristic beta of the SMB factor is similar to Table \ref{t2}: in the long run, the  smaller firm is   positively exposed and the larger firm is negatively exposed   to the SMB systematic risk. 

\subsection{Testing characteristic relevance in asset pricing models}

We now test the relevance of  characteristics on factor betas,  an important research question in asset pricing.     We consider the linear specification $\bg_{lt}=\bX_{lt}'\btheta_t$ for a common coefficient matrix $\btheta_t$ and test the relevance of each of the characteristics $\bX_{lt}=$ (size, value, momentum  and volatility) on the betas.  
Note that the  $(i,k)$-th element of $\btheta_t$, denoted by $\theta_{t,ik}$, represents the effect of characteristic $i$ on the $k$ the factor beta. Although this specifies a linear function $\bg_{t}(\cdot)$, $\bX_{lt}$ could include nonlinear (sieve) transformations of each characteristic.  Our test is uniformly valid over $\bgamma_{lt}$. To estimate $\btheta_t$, we use the linear sieve $\bPhi_t=(\bX_{1t},...,\bX_{pt})'$ and $\bP_t=\bPhi_t(\bPhi_t'\bPhi_t)^{-1}\bPhi_t'$.
Then
$$
\widehat\btheta_t=\left( \bPhi_t'\bPhi_t \right)^{-1}  \bPhi_t'\widehat \bG_t .
$$

We construct the bootstrap confidence intervals  for each component of the estimated $\btheta_t$ on each trading day, and  calculate the proportion of positive (and negative) significance each year. These results are reported in Table \ref{t4}. For most of the period, the volatility has  a  significantly positive effect on the market factor,  the value characteristic has a significantly positive effect on the HML factor,  and the size characteristic has a significantly negative effect on the  SMB factor. These results are consistent with the fitted $\bg_t(\cdot)$ functions in Figures \ref{fig:Gsize}-\ref{fig:Gvol} (in the appendix).  Also note that size has insignificant effects on the market beta. We explain this from two aspects: on one hand, the market beta is mostly affected by the volatility instrument, and once it is conditioned, the size is no longer significant. On the other hand, we focus on firms that constitute to the S\&P 500 index, whose sizes are relatively large, and are therefore not essential in explaining the market betas.

\begin{table}[htp]
\small
\caption{Proportion of significant characteristics}
\begin{center}
\begin{tabular}{c|ccccccccc}
\hline
\hline
 &   \multicolumn{4}{c}{positive significance} &&   \multicolumn{4}{c}{negative significance} \\
characteristics & Mkt & HML & SMB & RMW &  & Mkt & HML & SMB & RMW \\
\hline

   &   \multicolumn{9}{c}{2008}\\
size & 0.024 & 0.036 & 0.000 & 0.048 &  & 0.000 & 0.215 & 0.984 & 0.128 \\
value & 0.008 & 0.892 & 0.008 & 0.048 &  & 0.000 & 0.000 & 0.219 & 0.076 \\
momentum & 0.004 & 0.044 & 0.159 & 0.315 &  & 0.139 & 0.450 & 0.187 & 0.060 \\
volatility & 0.534 & 0.598 & 0.295 & 0.092 &  & 0.000 & 0.048 & 0.040 & 0.167 \\

   &   \multicolumn{9}{c}{2011}\\
size & 0.000 & 0.052 & 0.000 & 0.088 &  & 0.000 & 0.139 & 0.976 & 0.052 \\
value & 0.028 & 0.984 & 0.020 & 0.016 &  & 0.000 & 0.000 & 0.235 & 0.211 \\
momentum & 0.000 & 0.032 & 0.175 & 0.191 &  & 0.008 & 0.371 & 0.072 & 0.064 \\
volatility & 0.956 & 0.004 & 0.147 & 0.000 &  & 0.000 & 0.741 & 0.195 & 0.821 \\

   &   \multicolumn{9}{c}{2012}\\
size & 0.000 & 0.048 & 0.000 & 0.155 &  & 0.000 & 0.171 & 0.920 & 0.044 \\
value & 0.008 & 0.996 & 0.016 & 0.104 &  & 0.000 & 0.000 & 0.235 & 0.108 \\
momentum & 0.080 & 0.024 & 0.100 & 0.004 &  & 0.000 & 0.355 & 0.175 & 0.522 \\
volatility & 0.813 & 0.171 & 0.458 & 0.187 &  & 0.000 & 0.155 & 0.036 & 0.084 \\

\hline
\end{tabular}
\end{center}
\label{t4}
\end{table}%

 Finally, addtional numerical results are presented in  the appendix, where we plot the estimated $\bg_t(\cdot)$ functions fitted by B-splines.

\section{Conclusion}\label{sec: conclusion}

This paper studies a conditional factor model with a large number of individuals for high-frequency data.
     One of the key features  of our model is that we  specify the factor betas as functions of time-varying observed characteristics that pick up long-run beta movements   driven by structural changes in the economic environment and in firm- or industry-specific conditions,  plus a remaining  (idiosyncratic) component that captures high-frequency movements in beta, which    picks up short-run fluctuations in beta in periods of high market volatility.  
 The two components capture different aspects of market beta dynamics. We show that the model can be extended to a more general continuous-time many moment conditions setting, and estimated using two-step  GMM setting.

The limiting distribution of the estimated characteristic effect on the betas has a discontinuity when the strength of the idiosyncratic beta is near zero.  We provide a uniformly valid inference using a cross-sectional bootstrap procedure for the   characteristic betas, and do not need to pretest  to know whether or not the idiosyncratic beta exists, or their strengths.  
 
The proposed   framework can be extended for inference about general coefficients in  other  important econometric models, such as discrete time conditional factor models, panel data models with varying coefficients, and nonlinear panel models. In these models, the structural parameter may be decomposed into the sum of a characteristic driven component plus an orthogonal component. An important example would be linear panel data model with varying coefficients. In these models the asymptotic distribution of the estimated characteristic effect would also have a discontinuity, and thus standard ``plug-in" inference fails to hold uniformly. The proposed bootstrap framework would then be very useful in these models. We shall leave these studies for future research.
 
\small

\singlespacing

\bibliographystyle{ims}
\bibliography{liaoBib,yangBib}

\begin{thebibliography}{65}
\expandafter\ifx\csname natexlab\endcsname\relax\def\natexlab#1{#1}\fi
\expandafter\ifx\csname url\endcsname\relax
  \def\url#1{\texttt{#1}}\fi
\expandafter\ifx\csname urlprefix\endcsname\relax\def\urlprefix{URL }\fi

\bibitem[{Ait-Sahalia et~al.(2014)Ait-Sahalia, Kalnina and
  Xiu}]{ait2014idiosyncratic}
\textsc{Ait-Sahalia, Y.}, \textsc{Kalnina, I.} and \textsc{Xiu, D.} (2014).
\newblock The idiosyncratic volatility puzzle: A reassessment at high
  frequency.
\newblock Tech. rep., Working Paper, University of Chicago.

\bibitem[{A{\"\i}t-Sahalia and Xiu(2017)}]{ait2017using}
\textsc{A{\"\i}t-Sahalia, Y.} and \textsc{Xiu, D.} (2017).
\newblock Using principal component analysis to estimate a high dimensional
  factor model with high-frequency data.
\newblock \textit{Journal of Econometrics} \textbf{201} 384--399.

\bibitem[{Andrews(1999)}]{andrews1999estimation}
\textsc{Andrews, D.~W.} (1999).
\newblock Estimation when a parameter is on a boundary.
\newblock \textit{Econometrica} \textbf{67} 1341--1383.

\bibitem[{Andrews(2000)}]{andrews2000inconsistency}
\textsc{Andrews, D.~W.} (2000).
\newblock Inconsistency of the bootstrap when a parameter is on the boundary of
  the parameter space.
\newblock \textit{Econometrica} \textbf{68} 399--405.

\bibitem[{Andrews(2004)}]{andrews2004block}
\textsc{Andrews, D.~W.} (2004).
\newblock The block--block bootstrap: improved asymptotic refinements.
\newblock \textit{Econometrica} \textbf{72} 673--700.

\bibitem[{Andrews and Soares(2010)}]{andrews2010inference}
\textsc{Andrews, D.~W.} and \textsc{Soares, G.} (2010).
\newblock Inference for parameters defined by moment inequalities using
  generalized moment selection.
\newblock \textit{Econometrica} \textbf{78} 119--157.

\bibitem[{Ang et~al.(2009)Ang, Hodrick, Xing and Zhang}]{ang2009high}
\textsc{Ang, A.}, \textsc{Hodrick, R.~J.}, \textsc{Xing, Y.} and \textsc{Zhang,
  X.} (2009).
\newblock High idiosyncratic volatility and low returns: International and
  further us evidence.
\newblock \textit{Journal of Financial Economics} \textbf{91} 1--23.

\bibitem[{Ang and Kristensen(2012)}]{ang2012testing}
\textsc{Ang, A.} and \textsc{Kristensen, D.} (2012).
\newblock Testing conditional factor models.
\newblock \textit{Journal of Financial Economics} \textbf{106} 132--156.

\bibitem[{Avramov and Chordia(2006)}]{avramov2006asset}
\textsc{Avramov, D.} and \textsc{Chordia, T.} (2006).
\newblock Asset pricing models and financial market anomalies.
\newblock \textit{Review of Financial Studies} \textbf{19} 1001--1040.

\bibitem[{Bai and Liao(2017)}]{bai2017inferences}
\textsc{Bai, J.} and \textsc{Liao, Y.} (2017).
\newblock Inferences in panel data with interactive effects using large
  covariance matrices.
\newblock \textit{Journal of Econometrics} \textbf{200} 59--78.

\bibitem[{Bai and Ng(2002)}]{BN02}
\textsc{Bai, J.} and \textsc{Ng, S.} (2002).
\newblock Determining the number of factors in approximate factor models.
\newblock \textit{Econometrica} \textbf{70} 191--221.

\bibitem[{Barndorff-Nielsen et~al.(2008)Barndorff-Nielsen, Hansen, Lunde and
  Shephard}]{BN-HLN:2008}
\textsc{Barndorff-Nielsen, O.~E.}, \textsc{Hansen, P.~R.}, \textsc{Lunde, A.}
  and \textsc{Shephard, N.} (2008).
\newblock Designing realized kernels to measure the ex post variation of equity
  prices in the presence of noise.
\newblock \textit{Econometrica} \textbf{76} 1481--1536.

\bibitem[{Barndorff-Nielsen and Shephard(2004)}]{barndorff2004econometric}
\textsc{Barndorff-Nielsen, O.~E.} and \textsc{Shephard, N.} (2004).
\newblock Econometric analysis of realized covariation: High frequency based
  covariance, regression, and correlation in financial economics.
\newblock \textit{Econometrica} \textbf{72} 885--925.

\bibitem[{Bollerslev et~al.(2016)Bollerslev, Li and Todorov}]{BLT:2016}
\textsc{Bollerslev, T.}, \textsc{Li, S.~Z.} and \textsc{Todorov, V.} (2016).
\newblock Roughing up beta: Continuous vs. discontinuous betas, and the cross
  section of expected stock returns.
\newblock \textit{Journal of Financial Economics} \textbf{120} 464--490.

\bibitem[{Cai et~al.(2012)Cai, Yuan et~al.}]{cai2012adaptive}
\textsc{Cai, T.~T.}, \textsc{Yuan, M.} \textsc{et~al.} (2012).
\newblock Adaptive covariance matrix estimation through block thresholding.
\newblock \textit{The Annals of Statistics} \textbf{40} 2014--2042.

\bibitem[{Chamberlain and Rothschild(1983)}]{CR}
\textsc{Chamberlain, G.} and \textsc{Rothschild, M.} (1983).
\newblock Arbitrage, factor structure and mean-variance analyssi in large asset
  markets.
\newblock \textit{Econometrica} \textbf{51} 1305--1324.

\bibitem[{Chen and Liao(2015)}]{chen2015sieve}
\textsc{Chen, X.} and \textsc{Liao, Z.} (2015).
\newblock Sieve semiparametric two-step gmm under weak dependence.
\newblock \textit{Journal of Econometrics} \textbf{189} 163--186.

\bibitem[{Chen et~al.(2003)Chen, Linton and Van~Keilegom}]{chen2003estimation}
\textsc{Chen, X.}, \textsc{Linton, O.} and \textsc{Van~Keilegom, I.} (2003).
\newblock Estimation of semiparametric models when the criterion function is
  not smooth.
\newblock \textit{Econometrica}  1591--1608.

\bibitem[{Chen and Pouzo(2012)}]{chen2012estimation}
\textsc{Chen, X.} and \textsc{Pouzo, D.} (2012).
\newblock Estimation of nonparametric conditional moment models with possibly
  nonsmooth generalized residuals.
\newblock \textit{Econometrica} \textbf{80} 277--321.

\bibitem[{Chernozhukov et~al.(2016)Chernozhukov, Chetverikov, Demirer, Duflo,
  Hansen and Newey}]{chernozhukov2016double}
\textsc{Chernozhukov, V.}, \textsc{Chetverikov, D.}, \textsc{Demirer, M.},
  \textsc{Duflo, E.}, \textsc{Hansen, C.} and \textsc{Newey, W.~K.} (2016).
\newblock Double machine learning for treatment and causal parameters.
\newblock Tech. rep., cemmap working paper, Centre for Microdata Methods and
  Practice.

\bibitem[{Cochrane(1996)}]{cochrane1996cross}
\textsc{Cochrane, J.~H.} (1996).
\newblock A cross-sectional test of an investment-based asset pricing model.
\newblock \textit{Journal of Political Economy} \textbf{104} 572--621.

\bibitem[{Connor and Korajczyk(1993)}]{CK93}
\textsc{Connor, G.} and \textsc{Korajczyk, R.} (1993).
\newblock A test for the number of factors in an approximate factor model.
\newblock \textit{Journal of Finance} \textbf{48} 1263--1291.

\bibitem[{Connor and Korajczyk(1986)}]{connor1986performance}
\textsc{Connor, G.} and \textsc{Korajczyk, R.~A.} (1986).
\newblock Performance measurement with the arbitrage pricing theory: A new
  framework for analysis.
\newblock \textit{Journal of financial economics} \textbf{15} 373--394.

\bibitem[{Connor and Linton(2007)}]{CL07}
\textsc{Connor, G.} and \textsc{Linton, O.} (2007).
\newblock Semiparametric estimation of a characteristic-based factor model of
  stock returns.
\newblock \textit{Journal of Empirical Finance} \textbf{14} 694--717.

\bibitem[{Connor et~al.(2012)Connor, Matthias and Linton}]{CMO}
\textsc{Connor, G.}, \textsc{Matthias, H.} and \textsc{Linton, O.} (2012).
\newblock Efficient semiparametric estimation of the fama-french model and
  extensions.
\newblock \textit{Econometrica} \textbf{80} 713--754.

\bibitem[{Daniel and Titman(1997)}]{daniel1997evidence}
\textsc{Daniel, K.} and \textsc{Titman, S.} (1997).
\newblock Evidence on the characteristics of cross sectional variation in stock
  returns.
\newblock \textit{the Journal of Finance} \textbf{52} 1--33.

\bibitem[{Darolles et~al.(2011)Darolles, Fan, Florens and
  Renault}]{darolles2011nonparametric}
\textsc{Darolles, S.}, \textsc{Fan, Y.}, \textsc{Florens, J.-P.} and
  \textsc{Renault, E.} (2011).
\newblock Nonparametric instrumental regression.
\newblock \textit{Econometrica} \textbf{79} 1541--1565.

\bibitem[{Eilers and Marx(1996)}]{eilers1996flexible}
\textsc{Eilers, P.~H.} and \textsc{Marx, B.~D.} (1996).
\newblock Flexible smoothing with b-splines and penalties.
\newblock \textit{Statistical science}  89--102.

\bibitem[{Fama and French(1992)}]{FF}
\textsc{Fama, E.~F.} and \textsc{French, K.~R.} (1992).
\newblock The cross-section of expected stock returns.
\newblock \textit{Journal of Finance} \textbf{47} 427--465.

\bibitem[{Fan et~al.(2016{\natexlab{a}})Fan, Furger and
  Xiu}]{fan2016incorporating}
\textsc{Fan, J.}, \textsc{Furger, A.} and \textsc{Xiu, D.}
  (2016{\natexlab{a}}).
\newblock Incorporating global industrial classification standard into
  portfolio allocation: A simple factor-based large covariance matrix estimator
  with high-frequency data.
\newblock \textit{Journal of Business \& Economic Statistics} \textbf{34}
  489--503.

\bibitem[{Fan et~al.(2013)Fan, Liao and Mincheva}]{POET}
\textsc{Fan, J.}, \textsc{Liao, Y.} and \textsc{Mincheva, M.} (2013).
\newblock Large covariance estimation by thresholding principal orthogonal
  complements (with discussion).
\newblock \textit{Journal of the Royal Statistical Society, Series B}
  \textbf{75} 603--680.

\bibitem[{Fan et~al.(2016{\natexlab{b}})Fan, Liao and Wang}]{fan2016projected}
\textsc{Fan, J.}, \textsc{Liao, Y.} and \textsc{Wang, W.} (2016{\natexlab{b}}).
\newblock Projected principal component analysis in factor models.
\newblock \textit{Annals of statistics} \textbf{44} 219.

\bibitem[{Fan et~al.(2015)Fan, Liao and Yao}]{powerenhancement}
\textsc{Fan, J.}, \textsc{Liao, Y.} and \textsc{Yao, J.} (2015).
\newblock Power enhancement in high dimensional cross-sectional tests.
\newblock \textit{Econometrica} \textbf{83} 1497Ð1541.

\bibitem[{Ferson and Harvey(1999)}]{ferson1999conditioning}
\textsc{Ferson, W.~E.} and \textsc{Harvey, C.~R.} (1999).
\newblock Conditioning variables and the cross section of stock returns.
\newblock \textit{The Journal of Finance} \textbf{54} 1325--1360.

\bibitem[{Gagliardini et~al.(2016)Gagliardini, Ossola and
  Scaillet}]{gagliardini2016time}
\textsc{Gagliardini, P.}, \textsc{Ossola, E.} and \textsc{Scaillet, O.} (2016).
\newblock Time-varying risk premium in large cross-sectional equity data sets.
\newblock \textit{Econometrica} \textbf{84} 985--1046.

\bibitem[{Ghysels(1998)}]{ghysels1998stable}
\textsc{Ghysels, E.} (1998).
\newblock On stable factor structures in the pricing of risk: Do time-varying
  betas help or hurt?
\newblock \textit{The Journal of Finance} \textbf{53} 549--573.

\bibitem[{Hall and Horowitz(2005)}]{hall2005nonparametric}
\textsc{Hall, P.} and \textsc{Horowitz, J.~L.} (2005).
\newblock Nonparametric methods for inference in the presence of instrumental
  variables.
\newblock \textit{The Annals of Statistics} \textbf{33} 2904--2929.

\bibitem[{Herskovic et~al.(2016)Herskovic, Kelly, Lustig and
  Van~Nieuwerburgh}]{herskovic2016common}
\textsc{Herskovic, B.}, \textsc{Kelly, B.}, \textsc{Lustig, H.} and
  \textsc{Van~Nieuwerburgh, S.} (2016).
\newblock The common factor in idiosyncratic volatility: Quantitative asset
  pricing implications.
\newblock \textit{Journal of Financial Economics} \textbf{119} 249--283.

\bibitem[{Jacod et~al.(2009)Jacod, Li, Mykland, Podolskij and
  Vetter}]{Jacod&LMPV:2009}
\textsc{Jacod, J.}, \textsc{Li, Y.}, \textsc{Mykland, P.~A.},
  \textsc{Podolskij, M.} and \textsc{Vetter, M.} (2009).
\newblock Microstructure noise in the continuous case: The pre-averaging
  approach.
\newblock \textit{Stochastic Processes and their Applications} \textbf{119}
  2249--2276.
\newline\urlprefix\url{http://ideas.repec.org/a/eee/spapps/v119y2009i7p2249-2276.html}

\bibitem[{Jacod and Protter(2011)}]{Jacod&protter:2011}
\textsc{Jacod, J.} and \textsc{Protter, P.} (2011).
\newblock \textit{Discretization of Processes}.
\newblock Springer.

\bibitem[{Jacod and Rosenbaum(2013)}]{Jacod&Rosenbaum:2013}
\textsc{Jacod, J.} and \textsc{Rosenbaum, M.} (2013).
\newblock Quarticity and other functionals of volatility: Efficient estimation.
\newblock \textit{The Annals of Statistics} \textbf{41} 1462--1484.

\bibitem[{Jagannathan and Wang(1996)}]{jagannathan1996conditional}
\textsc{Jagannathan, R.} and \textsc{Wang, Z.} (1996).
\newblock The conditional capm and the cross-section of expected returns.
\newblock \textit{The Journal of finance} \textbf{51} 3--53.

\bibitem[{Kalnina(2012)}]{kalnina2012nonparametric}
\textsc{Kalnina, I.} (2012).
\newblock Nonparametric tests of time variation in betas.
\newblock \textit{Universit{\'e} de Montr{\'e}al} .

\bibitem[{Kelly et~al.(2017)Kelly, Pruitt and Su}]{kelly2017instrumented}
\textsc{Kelly, B.~T.}, \textsc{Pruitt, S.} and \textsc{Su, Y.} (2017).
\newblock Instrumented principal component analysis .

\bibitem[{Ketz(2017)}]{ketz2017testing}
\textsc{Ketz, P.} (2017).
\newblock Testing overidentifying restrictions when the true parameter vector
  is near or at the boundary of the parameter space.
\newblock Tech. rep., Working Paper.

\bibitem[{Ketz(2018)}]{ketz2018subvector}
\textsc{Ketz, P.} (2018).
\newblock Subvector inference when the true parameter vector may be near or at
  the boundary .

\bibitem[{Kim et~al.(2018)Kim, Korajczyk and Neuhierl}]{kim2018arbitrage}
\textsc{Kim, S.}, \textsc{Korajczyk, R.~A.} and \textsc{Neuhierl, A.} (2018).
\newblock Arbitrage portfolios in large panels.
\newblock \textit{Available at SSRN} .

\bibitem[{King et~al.(1994)King, Sentana and Wadhwani}]{king1994volatility}
\textsc{King, M.}, \textsc{Sentana, E.} and \textsc{Wadhwani, S.} (1994).
\newblock Volatility and links between national stock markets.
\newblock \textit{Econometrica} \textbf{62} 901--33.

\bibitem[{Lahiri(1999)}]{lahiri1999theoretical}
\textsc{Lahiri, S.~N.} (1999).
\newblock Theoretical comparisons of block bootstrap methods.
\newblock \textit{Annals of Statistics}  386--404.

\bibitem[{Lettau and Ludvigson(2001)}]{lettau2001resurrecting}
\textsc{Lettau, M.} and \textsc{Ludvigson, S.} (2001).
\newblock Resurrecting the (c) capm: A cross-sectional test when risk premia
  are time-varying.
\newblock \textit{Journal of political economy} \textbf{109} 1238--1287.

\bibitem[{Li et~al.(2016)Li, Todorov and Tauchen}]{li2016inference}
\textsc{Li, J.}, \textsc{Todorov, V.} and \textsc{Tauchen, G.} (2016).
\newblock Inference theory for volatility functional dependencies.
\newblock \textit{Journal of Econometrics} \textbf{193} 17--34.

\bibitem[{Li et~al.(2017)Li, Todorov and Tauchen}]{li2017adaptive}
\textsc{Li, J.}, \textsc{Todorov, V.} and \textsc{Tauchen, G.} (2017).
\newblock Adaptive estimation of continuous-time regression models using
  high-frequency data.
\newblock \textit{Journal of Econometrics} \textbf{200} 36--47.

\bibitem[{Li et~al.(2018)Li, Todorov and Tauchen}]{li2018jump}
\textsc{Li, J.}, \textsc{Todorov, V.} and \textsc{Tauchen, G.} (2018).
\newblock Jump factor models in large cross-sections.

\bibitem[{Li and Xiu(2016)}]{li2016generalized}
\textsc{Li, J.} and \textsc{Xiu, D.} (2016).
\newblock Generalized method of integrated moments for high-frequency data.
\newblock \textit{Econometrica} \textbf{84} 1613--1633.

\bibitem[{Mancini(2001)}]{Mancini:2001}
\textsc{Mancini, C.} (2001).
\newblock Disentangling the jumps of the diffusion in a geometric jumping
  {B}rownian motion.
\newblock \textit{Giornale dell'Istituto Italiano degli Attuari} \textbf{LXIV}
  19--41.

\bibitem[{Mykland et~al.(2006)Mykland, Zhang et~al.}]{mykland2006anova}
\textsc{Mykland, P.~A.}, \textsc{Zhang, L.} \textsc{et~al.} (2006).
\newblock Anova for diffusions and ito processes.
\newblock \textit{The Annals of Statistics} \textbf{34} 1931--1963.

\bibitem[{Newey(1994)}]{newey1994asymptotic}
\textsc{Newey, W.~K.} (1994).
\newblock The asymptotic variance of semiparametric estimators.
\newblock \textit{Econometrica}  1349--1382.

\bibitem[{Newey and Powell(2003)}]{newey2003instrumental}
\textsc{Newey, W.~K.} and \textsc{Powell, J.~L.} (2003).
\newblock Instrumental variable estimation of nonparametric models.
\newblock \textit{Econometrica} \textbf{71} 1565--1578.

\bibitem[{Pedersen(2017)}]{pedersen2017inference}
\textsc{Pedersen, R.~S.} (2017).
\newblock Inference and testing on the boundary in extended constant
  conditional correlation garch models.
\newblock \textit{Journal of Econometrics} \textbf{196} 23--36.

\bibitem[{Pelger(2016)}]{Pelger:2016}
\textsc{Pelger, M.} (2016).
\newblock Large-dimensional factor modeling based on high-frequency
  observations.
\newblock Working paper.

\bibitem[{Protter(2005)}]{Protter:2005}
\textsc{Protter, P.~E.} (2005).
\newblock \textit{Stochastic Integration and Differential Equations}.
\newblock 2nd ed. Springer.

\bibitem[{Shanken(1990)}]{shanken1990intertemporal}
\textsc{Shanken, J.} (1990).
\newblock Intertemporal asset pricing: An empirical investigation.
\newblock \textit{Journal of Econometrics} \textbf{45} 99--120.

\bibitem[{Stock and Watson(2002)}]{SW02}
\textsc{Stock, J.} and \textsc{Watson, M.} (2002).
\newblock Forecasting using principal components from a large number of
  predictors.
\newblock \textit{Journal of the American Statistical Association} \textbf{97}
  1167--1179.

\bibitem[{White(1980)}]{white1980heteroskedasticity}
\textsc{White, H.} (1980).
\newblock A heteroskedasticity-consistent covariance matrix estimator and a
  direct test for heteroskedasticity.
\newblock \textit{Econometrica}  817--838.

\bibitem[{Zhang et~al.(2005)Zhang, Mykland and A\"{i}t-Sahalia}]{Zhang&MA:2005}
\textsc{Zhang, L.}, \textsc{Mykland, P.~A.} and \textsc{A\"{i}t-Sahalia, Y.}
  (2005).
\newblock A tale of two time scales: Determining integrated volatility with
  noisy high-frequency data.
\newblock \textit{Journal of the American Statistical Association} \textbf{100}
  1394--1411.

\end{thebibliography}

\end{document}